\documentclass[preprint]{aastex}
\usepackage{amsmath}
\usepackage{xspace}

\def\ie{i.e.\xspace}
\def\eg{e.g.\xspace}
\def\ltsima{$\; \buildrel < \over \sim \;$}
\def\simlt{\lower.5ex\hbox{\ltsima}}
\def\gtsima{$\; \buildrel > \over \sim \;$}
\def\simgt{\lower.5ex\hbox{\gtsima}}
\def\fesc{{$\langle f_{esc}\rangle$}\xspace}

\def\H2{H$_2$\xspace} 
\def\m{$^{-1}$\xspace}
\def\mm{$^{-2}$\xspace}
\def\mmm{$^{-3}$\xspace}
\def\pp{$^2$\xspace}

\def\ion#1#2{\text{#1\,\sc #2}}
\def\HI{{\ion{H}{i} }}
\def\HII{{\ion{H}{ii} }}
\def\GI{{\ion{He}{i} }}
\def\GII{{\ion{He}{ii} }}
\def\GIII{{\ion{He}{iii} }}

\def\ju{erg cm\mm s\m Hz\m sr\m}
\def\sfru{M$_\odot$~yr\m Mpc\mmm}
\def\popII{Population~II\xspace}
\def\pop3{Population~III\xspace}
\def\pII{``large-halo''\xspace}
\def\p3{``small-halo''\xspace}
\def\pp3{``Small-halo''\xspace}
\def\Mpc{$h^{-1}$~Mpc\xspace}
\def\kpc{$h^{-1}$~kpc\xspace}
\def\pc{$h^{-1}$~pc\xspace}
\def\Ms{$h^{-1}$~M$_\odot$\xspace}
 
\received{.......}
\accepted{.......}
\slugcomment{to be submitted to ApJ }

\begin{document}
\tighten
\thispagestyle{empty}

\pagestyle{myheadings}
\markright{DRAFT: \today\hfill}

\def\placefig#1{#1}

\title{THE FATE OF THE FIRST GALAXIES. II.\\
EFFECTS OF RADIATIVE FEEDBACK}

\author{MASSIMO RICOTTI, NICKOLAY Y. GNEDIN, AND J. MICHAEL SHULL$^1$ } 

\affil{
Center for Astrophysics and Space Astronomy \\
Department of Astrophysical and Planetary Sciences \\
University of Colorado, Campus Box 389, Boulder CO 80309 \\
E--mail: ricotti, gnedin, mshull@casa.colorado.edu \\ 
$^1$ also at JILA, University of Colorado and National
Institute of Standards and Technology }  
 
\begin{abstract}
 
  We use 3D cosmological simulations with radiative transfer to study
  the formation and evolution of the first galaxies in a $\Lambda CDM$
  cosmology. The simulations include continuum radiative transfer
  using the ``Optically Thin Variable Eddington Tensor'' (OTVET)
  approximation and line-radiative transfer in the \H2 Lyman-Werner bands
  of the UV background radiation. Chemical and thermal processes are
  treated in detail, particularly the ones relevant for H$_2$
  formation and destruction.  
  
  We find that the first luminous objects (``small-halos'') are
  characterized by bursting star formation (SF) that is self-regulated
  by a feedback process acting on cosmological instead of galactic
  scales. The global star formation history is regulated by the mean
  number of ionizing photons that escape from each source,
  $\epsilon_{UV}$\fesc.  It is almost independent of the assumed star
  formation efficiency parameter, $\epsilon_*$, and the intensity of
  the dissociating background. The main feedback process that
  regulates the SF is the re-formation of H$_2$ in front of \HII
  regions and inside relic \HII regions. The \HII regions remain
  confined inside filaments, maximizing the production of H$_2$ in
  overdense regions through cyclic destruction/reformation of H$_2$.
  If $\epsilon_{UV}$\fesc$ > 10^{-7}/\epsilon_*$ the SF is
  self-regulated, photo-evaporation of \p3 objects dominate the metal
  pollution of the low density IGM, and the mass of produced metals
  depends only on \fesc. If $\epsilon_{UV}$\fesc$ \simlt
  10^{-7}/\epsilon_*$, positive feedback dominates, and \p3 objects
  constitute the bulk of the mass in stars and metals at least until
  redshift $z \sim 10$. \pp3 objects cannot reionize the universe
  because the feedback mechanism confines the \HII regions inside the
  large scale structure filaments.  In contrast to massive objects
  (``large halos''), which can reionize voids, \p3 objects partially
  ionize only the dense filaments while leaving the voids mostly
  neutral.

\end{abstract}

\section{Introduction \label{sec:int}}

In cold dark matter (CDM) cosmologies, \p3 galaxies are believed to be
the first luminous objects formed in the universe. Defined as small
mass protogalaxies, \p3 objects have virial temperature $T_{vir} <
10^4$ K and rely on H$_2$ line cooling to form stars. The first
generation of stars is necessarily metal-free, as all elements heavier
than Li are produced in the cores of stars or by supernova (SN)
explosions.  In the literature \pop3 is often used to refer to both
metal-free stars and to protogalaxies with $T_{vir}< 10^4$~K. In order
to avoid this confusion, we use the term \p3 objects instead of the
widely used ``\pop3 objects'' throughout our paper.  For instance, the
stars in a \p3 object are not necessarily \pop3 because the
interstellar medium (ISM) could be polluted by metals rather quickly.
In this paper, we use synthetic stellar energy distributions (SEDs)
calculated for metal-free stars, which for brevity we call \pop3 SEDs,
and SEDs typical of low-metallicity stars, which we call \popII SEDs.
Although the SED does not change the star formation (hereafter SF)
history, we find that the \H2 abundance in the intergalactic medium
(IGM) depends on it.

In this paper, the second in a series on the formation and evolution
of the first galaxies, we focus on the mechanisms responsible for the
the self-regulation of star formation in \p3 objects.  In
\cite*{RicottiGSa:02} (Paper~I) we discussed the numerical methods and
the physics included in the simulations used in this paper. From
resolution studies we found that our largest simulations seem to be
close to numerical convergence. These models are the first 3D
cosmological simulations to include the physics necessary to study the
formation of \p3 objects and their radiative feedback
self-consistently. Mechanical feedback from SN explosions and stellar
winds is not included, as it is the subject of a separate paper. The
main new ingredients are the inclusion of continuum radiative transfer
and line radiative transfer in the Lyman-Werner bands. A number of
potentially relevant physical processes are included as well:
secondary ionizations of H and He, detailed \H2 chemistry and cooling
processes, heating by Ly$\alpha$ resonant scattering, H and He
recombination lines, metal production, and radiative cooling. The SED
of the sources is consistent with the choice of the escape fraction of
ionizing radiation, \fesc.  The results of our simulations are
considerably different from previous models of the formation of \p3
objects. These previous works were semi-analytic treatments
\citep*{HaimanRL:97, HaimanAR:00, CiardiF:00} or 3D simulations
\citep*{Machacek:00} without radiative transfer. In contrast to
previous studies, we find that the formation of \p3 objects is not
suppressed by the \H2 dissociating background.  The reason for this
difference is that these earlier studies did not include a primary
mechanism of positive feedback for the formation of \p3 objects.

Prior to our work, it was widely thought that the main feedback
mechanism that regulates \p3 formation is the build-up of the \H2
dissociating background. Indeed, the dissociating background can
suppress or delay the formation of \p3 objects if \H2 is not reformed
efficiently.  Because of these results, \p3 objects were believed to
be unimportant for subsequent cosmic evolution, metal enrichment of
the IGM, and reionization.  However, our recent work
\citep*{RicottiGS:01} pointed out a new radiative feedback mechanism
that turns out to provide the dominant regulation for the formation of
\p3 objects. A large amount of \H2 is naturally reformed in shells
(PFRs: positive feedback regions) in front of \HII regions or inside
recombined fossil \HII regions.  The \H2 formation rate is
proportional to the density squared of the gas and is very efficient
in the filaments and inside galaxies. The bursting mode of SF observed
in simulations produces fossil \HII regions and therefore allows \H2
to be continuously reformed.

Contrary to the initial ideas in \cite{RicottiGS:01}, the volume
filling factor of positive feedback regions (PFRs) has to remain small
to maximize the positive feedback.  Therefore, we do not require
ionizing escape fraction \fesc$\sim 1$, to maximize the IGM volume
occupied by PFRs with respect to the volume occupied by the
dissociating spheres (or with respect to the intensity of the
dissociating background). Instead, \p3 formation is possible if
\fesc$<1$; inside the dense filaments, the positive feedback of \H2
reformation always dominates the negative feedback of the dissociating
background and the of local dissociating radiation.  This, together
with the clustering of dark matter (DM) halos in the overdense
regions, maintains a sufficiently high \H2 abundance inside the
filaments to allow \p3 objects to cool and form stars.  On the
other hand, if the star formation rate (SFR) is too high (and
\fesc$\sim 1$), the filaments become highly ionized by the Str\"omgren
spheres and the \H2 is destroyed by direct ionization.  This produce a
local and temporary halt of SF until the Str\"omgren spheres
recombine.  Shells of \H2 then reform inside the \HII regions and a
new burst of SF occurs.
      
In \cite{RicottiGS:01} we emphasized the importance of PFRs in
reducing the dissociating background. In some simulations, the
production of \H2 is high enough to reduce the dissociating background
intensity by a factor of 10. But this does not affect the SF history.
In the same paper, we pointed out that enhanced galaxy formation is
possible if it happens inside a PFR.  Indeed, in this work we find
that this is the most important mechanism that regulates the formation
of \p3 objects.  Another positive feedback mechanism proposed by
\cite{Ferrara:98} involves local production of \H2 in shells produced
by SN explosions.  In the simulations presented in this paper, we do
not include feedback from SN explosions, although we plan to address
this problem in a subsequent paper.

In \cite*{RicottiS:00} we studied \fesc for \p3 objects, assuming a
spherical halo geometry.  The high merger rate in the early universe
should favor the formation of spheroidal galaxies instead of disks. We
have found that \fesc is small at high redshift, \fesc$\propto
\exp[-(1+z_{vir})]$.  This supports the theory that \p3 objects formed
copiously in the early universe, and their existence could be directly
or indirectly observed today. Metals in the Ly$\alpha$ forest and
dwarf spheroidal galaxies in the Local Group are two examples of
observations that can test our models. In Paper~III, currently in
preparation, we will study the properties of \p3 galaxies and try to
make a connection with available observational data on the dwarf
spheroidal galaxies in the Local Group.

The paper is organized in the following manner. In \S~\ref{sec:code}
we briefly review the physics included in the code and the free
parameters. This section serves as a quick reference of arguments
treated extensively in Paper~I .  The results of the simulations are
shown and discussed in \S~\ref{sec:res}. In \S~\ref{sec:sum} we
provide a summary and final comments.

\section{The Code\label{sec:code}}

The simulations were performed with the ``Softened Lagrangian
Hydrodynamics'' (SLH-P$^3$M) code described in detail in
\cite*{Gnedin:95, Gnedin:96a}, and \cite*{GnedinB:96}. The simulation
evolves collisionless DM particles, gas particles, ``star-particles''
formed using the Schmidt law in resolution elements that sink below
the numerical resolution of the code, and radiation, whose transfer is
treated self-consistently with the OTVET approximation of
\cite{GnedinA:01}. We also include line radiative transfer in the \H2
Lyman-Werner bands of the background radiation, secondary ionization
of H and He, heating by Ly$\alpha$ scattering, detailed \H2 chemistry
and cooling, and self-consistent SED of the sources (Paper~I).

We adopt a $\Lambda$CDM cosmological model with parameters: $\Omega_0
= 0.3$, $\Omega_\Lambda= 0.7$, $h=0.7$ and $\Omega_b = 0.04$.  The
initial spectrum of perturbations has $\sigma_8=0.91$ and $n=1$.  All
simulations start at $z = 100$ and finish at $z \simlt 9$. We use box
sizes $L_{box}=0.5, 1$ and 2 comoving \Mpc and grids with
$N_{box}^3=256^3, 128^3$ and $64^3$ cells. We achieve the maximum mass
resolution of $M_{DM}=4.93\times 10^3$ \Ms and spatial resolution of
$156$ comoving \pc in our biggest run.  We fully resolve the SF in
objects within the mass range $5 \times 10^5~{\rm M}_\odot \simlt
M_{DM} \simlt 10^9~{\rm M}_\odot$.

In Paper~I we discussed extensively the details of the code and the
physics included in the simulation. We also studied the numerical
convergence of the simulations that is especially crucial in the study
of the first objects. High mass resolution is needed because the
objects that we want to resolve have small masses ($10^5~{\rm M}_\odot
\simlt M_{DM} \simlt 10^8~{\rm M}_\odot$). Moreover, the box size has
to be large enough in order to include at least a few of the rare
first objects. The first \p3 objects should form at $z~\sim 30$ from
$3\sigma$ density perturbations; the first \pII objects (with $M_{DM}
\sim 5 \times 10^7$ \Ms) should form at $z \sim 20$, also from
$3\sigma$ perturbations.

The reader interested in numerical issues or in the details of the
physics included in the simulation should refer to Paper~I. Below, we
summarize the meaning of the four free parameters in the simulations.
\begin{itemize}
\item $\epsilon_*$: Star formation efficiency in the Schmidt law
  ($d\rho_*/dt = \epsilon_* \rho_g / t_*$, where $\rho_*$ and $\rho_g$
  are the stellar and gas density, respectively, and $t_*$ is the maximum
  between the dynamical and cooling time).
\item $\epsilon_{UV}$: Energy in ionizing photons per rest mass energy
  of H atoms ($m_H c^2$) transformed into stars. This parameter
  depends on the IMF and stellar metallicity.
\item \fesc: escape fraction of ionizing photons from the resolution element.
\item $g_\nu$: Normalized stellar energy distribution (SED). We use
  \pop3 and \popII SEDs with Salpeter ($1 < {\rm M}_* <
  100$~M$_\odot$) initial mass function (IMF), and we modify $g_\nu$
  according to the value of \fesc.\footnote{A small value of \fesc
    increases the flux jump at the Lyman limit and makes the SED
    harder.}
\end{itemize}

\section{Results\label{sec:res}}

In this section, we discuss the main physical processes that regulate
the formation and evolution of \p3 objects. We ran a large set ($\sim
20$) of simulations in order to explore the dependence of the results on
free parameters in the simulation.  We ran the simulations on the
Origins2000 supercomputer at the NCSA in
Urbana-Champaign, IL. The typical clock time to run a $64^3$,
$128^3$, and $256^3$ simulation from $z=100$ to $z=10$ is about $50,
1000$, and $> 23,000$ hours. The total computational time used to run
the simulations presented in this work is about $44,000$ hours.
In Table~\ref{tab:2}, as a quick reference, we list the simulations
with radiative transfer discussed in this section. Each simulation is
named, 64L05p2noLW+1 for example, using the following convention:
\begin{align*}
\underbrace{64}\underbrace{\text{L05}}&[\underbrace{\text{p2}}]
[\underbrace{\text{noLW}}][\underbrace{+1}]\\
N_{box}~\hookleftarrow~~~~/~~~&~\downarrow~~~~~~\backslash~~~~~~\hookrightarrow~\propto
\log (\epsilon_{UV}\langle f_{esc} \rangle)\\
L_{box}~~&\text{Pop}~~\text{comment}
\end{align*}
where $N_{box}^3$ is the number of cells in the box, $L_{box}$ is the
comoving size of the box in \Mpc, ``p2'' refers to \popII (metallicity
$Z =0.04~Z_\odot$) and ``p3'' to \pop3 (metal-free) stars. The brackets
indicate the optional part of the name.  The comment ``noLW''
indicates that the sources emit no Lyman-Werner \H2 photons, ``noRAD''
is used for simulations without radiative transfer, and ``noC''
indicates that radiative transfer and \H2 cooling are not included in
the simulation. The second optional part of the name is
$\log[(\epsilon_{UV}/4\pi)\langle f_{esc} \rangle /1.6 \times 10^{-5}$].
Values of zero are omitted from the name. We will attempt to make
clear the parameters of each simulation in the text. Therefore, a
reader does not need to memorize the notation or repeatedly refer to
Table~\ref{tab:2}.
 
\def\tabone{
\begin{deluxetable}{lccccccccc}
\tablecaption{List of simulations with radiative transfer.\label{tab:2}}
\tablewidth{0pt}
\tablecolumns{10}
\tableheadfrac{.3}
\tablehead{
\colhead{RUN} &\colhead{$N_{box}$} & \colhead{$L_{box}$} & \colhead{Mass
  Res.} & \colhead{$B_*$} & \colhead{$g_\nu$} & \colhead{\fesc} &
\colhead{$\left({\epsilon_{UV} \over 4\pi}\right)$\fesc} & \colhead{$\epsilon_*$} &
\colhead{$\left({\epsilon_{UV} \over 4\pi}\right)$\fesc$\epsilon_*$} \\
\colhead{} & \colhead{} & \colhead{($h^{-1}$ Mpc)} & \colhead{($h^{-1}$ M$_\odot$
)} & \colhead{} & \colhead{} & \colhead{} & \colhead{} & \colhead{} &
\colhead{}
 }
\startdata
64L05noRAD & 64  & 0.5  & $3.94\times 10^4$ & 10 & - & 1 & 0 & 0.2 & 0\\
64L05noC   & 64  & 0.5  & $3.94\times 10^4$ & 10 & - & 1 & 0 & 0.2 & 0\\

64L05p2noLW-2 & 64  & 0.5 & $3.94\times 10^4$ & 10 & II & 1 & $1.6\times 10^{-7}$ & 0.2 & $3.2\times 10^{-8}$\\
64L05p2noLW   & 64  & 0.5 & $3.94\times 10^4$ & 10 & II & 1 & $1.6\times 10^{-5}$ & 0.2 & $3.2\times 10^{-6}$ \\
64L05p2noLW+1 & 64  & 0.5 & $3.94\times 10^4$ & 10 & II & 1 & $1.6\times 10^{-4}$ & 0.02 & $3.2\times 10^{-6}$ \\

64L05p2-2  & 64  & 0.5  & $3.94\times 10^4$ & 10 & II  & 1 & $1.6\times 10^{-7}$ & 0.2& $3.2\times 10^{-8}$ \\
64L05p2-1  & 64  & 0.5  & $3.94\times 10^4$ & 10 & II  & 1 & $1.6\times 10^{-6}$ & 0.2 & $3.2\times 10^{-7}$ \\
64L05p2    & 64  & 0.5  & $3.94\times 10^4$ & 10 & II  & 1 & $1.6\times 10^{-5}$ & 0.2 & $3.2\times 10^{-6}$ \\

64L05p3  & 64  & 0.5  & $3.94\times 10^4$ & 10 & III & 1 & $1.6\times 10^{-5}$ & 0.2 & $3.2\times 10^{-6}$ \\
64L05p3b & 64  & 0.5  & $3.94\times 10^4$ & 10 & III & 1 & $1.6\times 10^{-5}$ & 0.02& $3.2\times 10^{-7}$ \\
64L05p3c & 64  & 0.5  & $3.94\times 10^4$ & 10 & III & 1 & $1.6\times 10^{-5}$ & 0.002& $3.2\times 10^{-8}$ \\

64L05p2-1f & 64 & 0.5  & $3.94\times 10^4$ & 10 &
II\tablenotemark{a} & 0.1 & $1.6\times 10^{-5}$ & 0.2 &
$3.2\times 10^{-7}$\\
64L05p2-2f & 64 & 0.5  & $3.94\times 10^4$ & 10 &
II\tablenotemark{a} & $10^{-2}$ & $1.6\times 10^{-5}$ & 0.2 &
$3.2\times 10^{-8}$\\
64L05p3-2f & 64 & 0.5  & $3.94\times 10^4$ & 10 &
III\tablenotemark{b} & $10^{-2}$ & $1.6\times 10^{-5}$ & 0.2 &
$3.2\times 10^{-8}$\\
64L05p2-2fa & 64 & 0.5  & $3.94\times 10^4$ & 10 &
II\tablenotemark{a} & $10^{-2}$ & $1.6\times 10^{-5}$ & 0.05 &
$8\times 10^{-9}$\\
64L05p3-2fa & 64 & 0.5  & $3.94\times 10^4$ & 10 &
III\tablenotemark{b} & $10^{-2}$ & $1.6\times 10^{-5}$ & 0.05 &
$8\times 10^{-9}$\\
64L05p2-3 & 64 & 0.5  & $3.94\times 10^4$ & 10 &
II\tablenotemark{a} & $10^{-3}$ & $1.6\times 10^{-5}$ & 0.2 &
$3.2\times 10^{-9}$\\
64L05p2-5 & 64 & 0.5  & $3.94\times 10^4$ & 10 &
II\tablenotemark{a} & $10^{-5}$ & $1.6\times 10^{-5}$ & 0.2 &
$3.2\times 10^{-11}$\\

64L05p3b-3n & 64 & 0.5  & $3.94\times 10^4$ & 10 & III & 1 & $1.6\times
10^{-8}$ & 0.02 & $3.2\times 10^{-10}$\\
64L05p3b-3\tablenotemark{c}  & 64 & 0.5  & $3.94\times 10^4$ & 10 &
III\tablenotemark{b} & $10^{-3}$ & $1.6\times 10^{-8}$ & 0.02 &
$3.2\times 10^{-10}$\\[5pt]

64L1noRAD   & 64  & 1.0  & $3.15\times 10^5$ & 10 & -  & 1 & 0 & 0.2 & 0\\
64L1noC     & 64  & 1.0  & $3.15\times 10^5$ & 10 & -  & 1 & 0 & 0.2 & 0\\
64L1p2      & 64  & 1.0  & $3.15\times 10^5$ & 10 & II  & 1 & $1.6\times 10^{-5}$ & 0.2 & $3.2\times 10^{-6}$ \\
\\
128L05noRAD & 128 & 0.5  & $4.93\times 10^3$ & 10 & - & 1 & 0 & 0.2 & 0\\
128L05p2    & 128 & 0.5  & $4.93\times 10^3$ & 10 & II  & 1 & $1.6\times 10^{-5}$ & 0.2 & $3.2\times 10^{-6}$\\[5pt]

128L1noRAD   & 128 & 1.0  & $3.94\times 10^4$ & 10 & - & 1 & 0 & 0.2 & 0\\
128L1p2      & 128 & 1.0  & $3.94\times 10^4$ & 10 & II  & 1 & $1.6\times 10^{-5}$ &0.2 & $3.2\times 10^{-6}$\\
128L1p2-2\tablenotemark{c} & 128 & 1.0  & $3.94\times 10^4$ & 16 & II\tablenotemark{a} & $10^{-2}$ & $1.1\times 10^{-7}$ & 0.05 & $5.5\times 10^{-9}$\\
\\
256L1noRAD & 256 & 1.0  & $4.93\times 10^3$ & 10 & - & 1 & 0 & 0.2 & 0\\
256L1p3\tablenotemark{c}   & 256 & 1.0  & $4.93\times 10^3$ & 25 & III\tablenotemark{b} & 0.1 & $2.5\times 10^{-6}$ & 0.1 & $2.5\times 10^{-7}$\\
 \enddata 
\tablecomments{Parameter description. {\em Numerical parameters:}
   $N_{box}^3$ is the number of grid cells, $L_{box}$ is the box size
   in comoving \Mpc, $B_*$ is the parameter that regulates the maximum
   deformation of the Lagrangian mesh: the spatial resolution is $\sim
   L_{box}/(N_{box}B_*)$. {\em Physical parameters:} $g_\nu$ is the
   normalized SED (II=\popII and III=\pop3), $\epsilon_*$ is the star
   formation efficiency, $\epsilon_{UV}$ is the ratio of energy
   density of the ionizing radiation field to the gas rest-mass energy
   density converted into stars (depends on the IMF), and \fesc is the
   escape fraction of ionizing photons from the resolution element.}
\tablenotetext{a}{$g_\nu$ is modified assuming
   $a_0=N_{\GI}/N_{\HI}=0.1$ and $a_1=N_{\GII}/N_{\HI}=10$ where
   $N_i$ is the column density of the species/ion $i$ (see Paper~I).} 
\tablenotetext{b}{$g_\nu$ is modified assuming $a_0=0.01$, $a_1=10$.} 
\tablenotetext{c}{Secondary ionizations included.}
\end{deluxetable}
}
\placefig{\tabone}

In Figure~\ref{fig:r1}~(left) we show the comoving SFR (\sfru) for the
64L05noRAD, 64L05noC, 64L05p2, 128L05noRAD, and 128L05p2 runs. These
simulations have $L_{box}=0.5$ \Mpc and $\epsilon_*=0.2$. The two
solid lines show the SFR in the $128^3$ box (thick line) and in the
$64^3$ box (thin line) without radiative transfer. In the higher mass
resolution simulation, the SFR is larger since more small mass objects
are formed.  The short-dashed line shows the $64^3$ simulation without
radiative transfer and \H2 cooling. By definition, this simulation
does not form any \p3 objects. Comparing the short-dashed line with
the solid lines, we see that \p3 objects dominate the SFR down to $z
\sim 10$ if we do not include radiative feedback. The two long-dashed
lines show the SFR including radiative transfer (using the \popII SED
and $(\epsilon_{UV}/4\pi)$\fesc$=1.6 \times 10^{-5}$). Again, the thick line
is for the $128^3$ box and the thin line is the $64^3$ box. From
Figure~\ref{fig:r1} it is clear that SF is bursting and is
suppressed with respect to the simulations without radiative transfer.

\def\capfiga{(Left) SFR in 0.5 \Mpc box simulations: 64L05noRAD,
  64L05noC, 64L05p2, 128L05noRAD, and 128L05p2. The solid lines show
  simulations without radiative transfer and the long-dashed lines
  including radiative transfer. The short-dashed line shows a simulation
  without either radiative transfer or H$_2$ cooling. (Right) The
  same as on the left figure but for 1 \Mpc box simulations: 64L1noRAD,
  64L1noC, 64L1p2, 128L1noRAD, and 128L1p2. The 64L1 simulations have
  insufficient mass resolution to form \p3 objects.}
\placefig{
\begin{figure*}[thp]
\plottwo{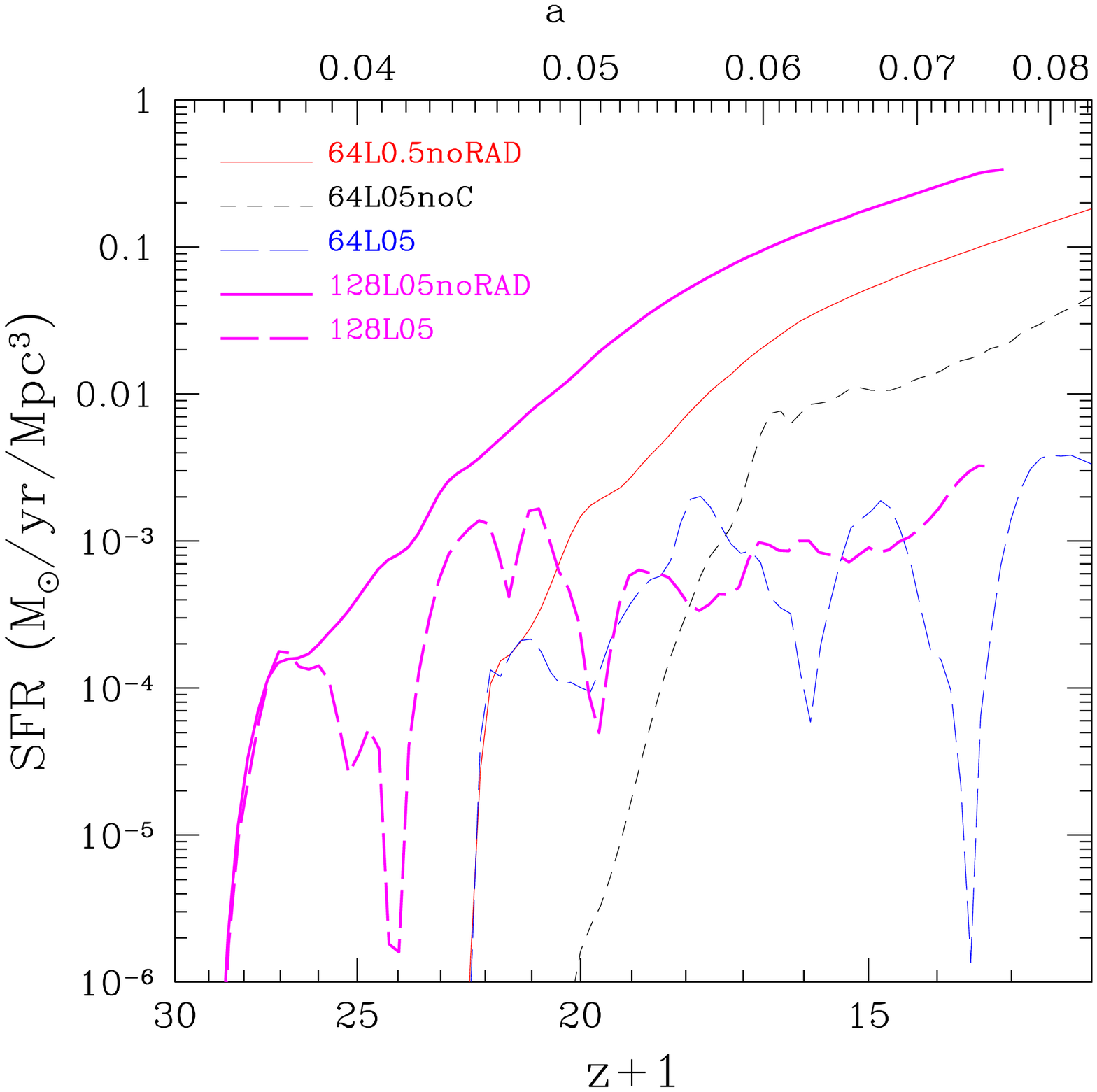}{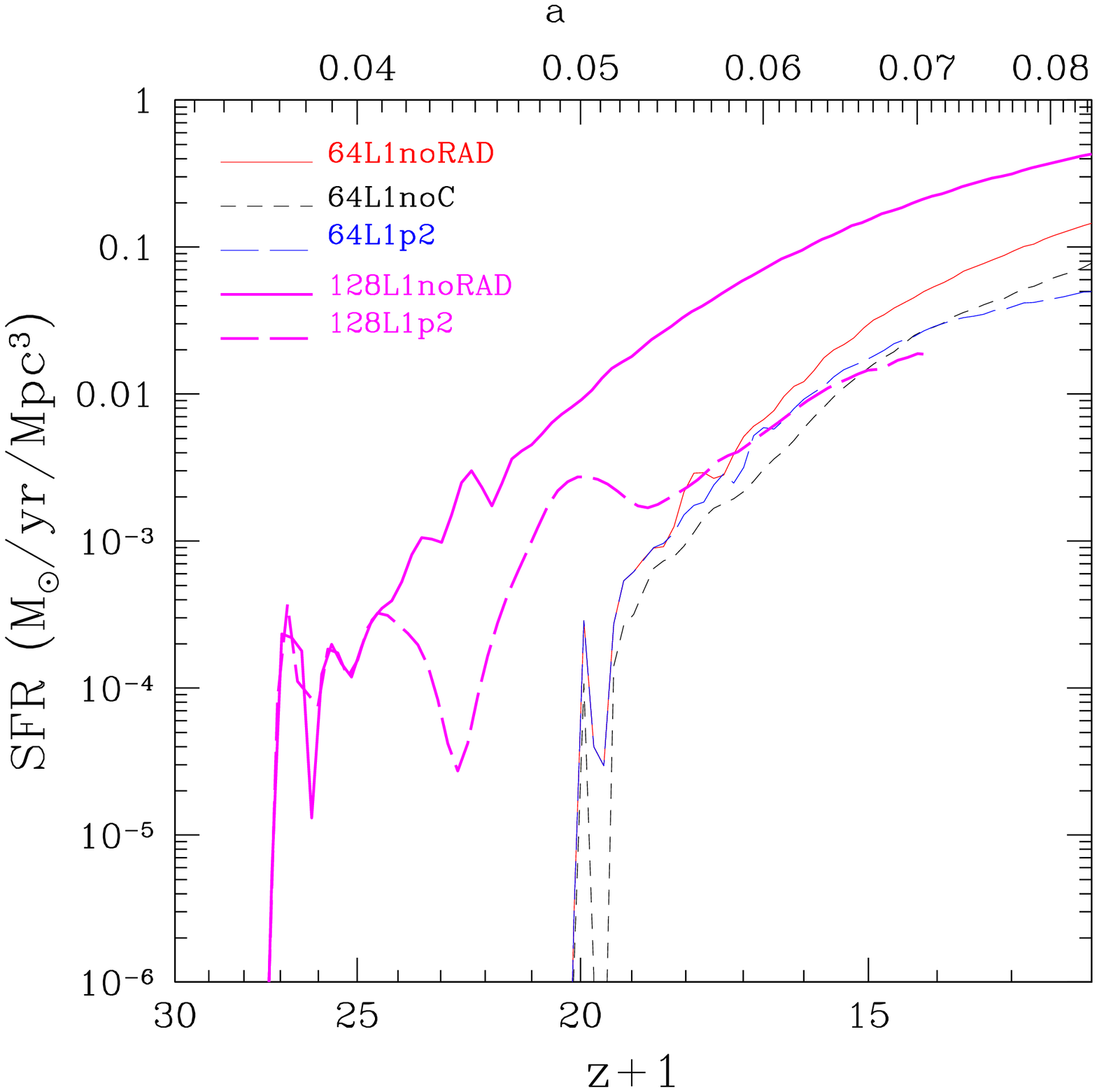}
\caption{\label{fig:r1}\capfiga}
\end{figure*}
}

Figure~\ref{fig:r1}~(right) shows the analogous simulations to
Figure~\ref{fig:r1}~(left) but for box size $L_{box}=1$ \Mpc. The
three thin lines show the SFR for the $64^3$ box: without radiative
transfer (solid), without radiative transfer and \H2 cooling (dashed),
and with radiative transfer (long-dashed). The lines are almost
indistinguishable, showing that we are forming only \pII objects and
that radiative feedback has no effect on their SFR. The DM mass
resolution of these simulations, $M_{DM} = 3.15\times 10^5$ \Ms, is
not sufficient to fully resolve the first \p3 objects, with typical
masses $M_{DM} \sim 10^6-10^8$ \Ms. In Paper~I we showed that we need
to resolve each object with at least 100 DM particles.  The two thick
lines show the SFR for the $128^3$ boxes without radiative transfer
(solid line) and including radiative transfer (long-dashed line). For
the simulation with radiative transfer we use the \popII SED and
$(\epsilon_{UV}/4\pi)$\fesc$=1.6 \times 10^{-5}$. This simulation has mass
resolution $M_{DM}=3.94 \times 10^4$ \Ms, sufficient to resolve \p3
objects, and box size large enough to include the first \pII objects
that form at $z \sim 20$. It is shown clearly that the star formation
is bursting at $z \simlt 20$ when it is dominated by \p3 objects. In
this simulation, \pII objects dominate the SFR as soon as they form,
producing the continuous star formation mode observed at $z \simgt
20$.

In summary, simulations with mass resolution $M_{DM} < 3.15\times
10^5$ \Ms (such as the 64L1 runs) do not form \p3 objects and
therefore are not well-suited for this study. On the other hand, if
the box is much smaller than 1 \Mpc (comoving), the formation of the
first \pII objects is delayed considerably. If $L_{box} \ge 1$ \Mpc,
the first rare massive objects form at $z \sim 20$. From the results
of other simulations, not shown here, we have verified that the global
SFR does not change if we increase the box size from $L_{box}=1$ \Mpc
to $L_{box}=8$ \Mpc, keeping the mass resolution constant (for
redshifts $z > 4$).  The bursting SF mode of \p3 objects is not
synchronized throughout the whole universe. Therefore, the strong
oscillations of the SFR observed in some of our simulations are an
artifact of the finite (actually quite small) size of the simulation
box.

In Figure~\ref{fig:tile} we show a sequence of four 3D-views of the
cube for the 128L1p2 run at $z=21.2, 17.2, 15.7, 13.3$. The rendering
of the volume is obtained by assigning quadratic opacity to the
logarithm of the gas density and linear opacity to the $x_{\HII}$
fraction. The colors show the $x_{\HII}$ fraction. The cosmological
\HII regions expand quickly in the IGM, but when their size becomes on
the order of the filamentary structure the expansion stops (in
comoving coordinates). The filaments become partially ionized, but the
voids remain neutral and reionization cannot occur. At redshift $z
\sim 20$ the first \pII objects form. For \pII objects, there is no
feedback process that can stop the \HII region from expanding into the
low density IGM. Indeed, these \HII regions never stop expanding and
eventually reionize the universe. Figure~\ref{fig:tile1} is analogous
to the previous figure, but here we show the \H2 abundance for the
64L05p3 run. The \H2 is quickly destroyed in the low density IGM by
the build-up of the dissociating background, but new \H2 is
continuously reformed in the filaments. Therefore, in the dense
regions where galaxy formation occurs, positive feedback dominates
over the negative feedback of the dissociating background. In the
following sections, we will show that the main mechanisms that
regulate galaxy formation of \p3 objects are the two positive feedback
processes found by \cite{RicottiGS:01}. In that paper, we discussed
the importance of \H2 shells (PFRs: positive feedback regions) that
form just in front each Str\"omgren sphere and inside relic
(recombining) \HII regions.

\def\capfigb{ Rendering of the 128L1p2 run at $z=21.2, 17.2, 15.7,
  13.3$. The colors code the logarithm of the \HII abundance. \pp3
  objects produce \HII regions that remain confined in the denser
  filaments, creating structures that resemble a pearl necklace. At
  $z =13.3$ the first \pII object produces a large \HII region
  (big blob on the left of the image) that continues to expand and
  eventually reionizes the IGM around it.}  
\placefig{
\begin{figure*}[thp]
\plotone{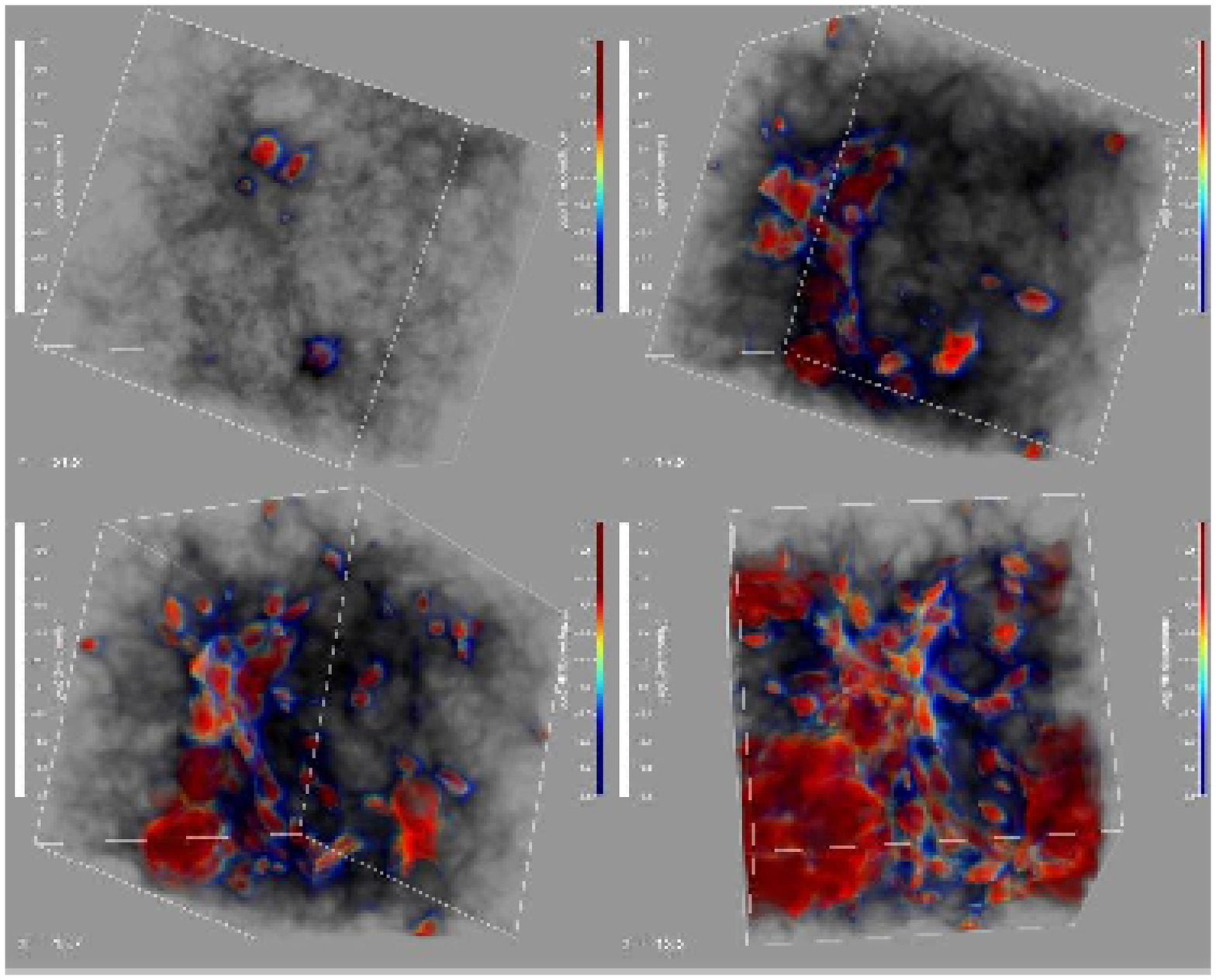}
\caption{\label{fig:tile}\capfigb}
\end{figure*}
} 
\def\capfigc{ Rendering of the 64L05p3 run at $z=15.7, 13.3, 12.3,
10.2$. The colors code the logarithm of the \H2 abundance.}
\placefig{
\begin{figure*}[thp]
\plotone{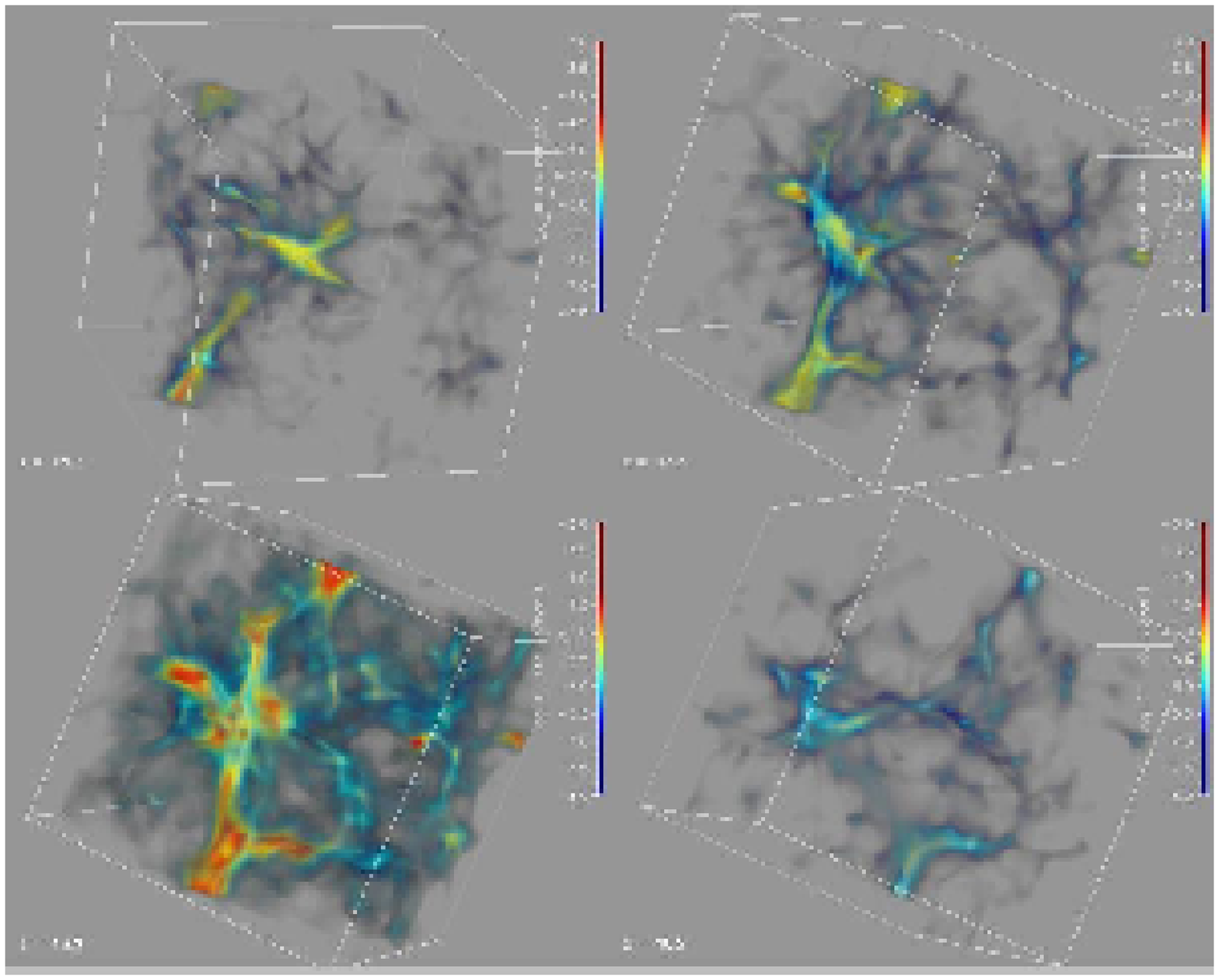}
\caption{\label{fig:tile1}\capfigc}
\end{figure*}
} 

Since the $128^3$ and $256^3$ cubes are computationally intensive, in
the following three sections (\S~\ref{ssec:LW}--\S~\ref{ssec:met})
we will try to understand the feedback mechanism that regulates the SF
of \p3 objects using $64^3$ cubes with $L_{box}=0.5$ \Mpc. We are
aware, though, that after $z \sim 20$ we are not properly including
the formation of the first \pII objects.  In \S~\ref{ssec:realsim} we
present the results of our largest simulations that include detailed
physics, and realistic values of the free parameters.

\subsection{Negative and Positive Feedback\label{ssec:LW}}

The background intensity in the Lyman-Werner bands determines the
redshift at which H$_2$ is destroyed in the low density IGM. In
Figure~\ref{fig:r2}~(top) we show the mean mass- and volume-weighted
molecular abundances $\langle x_{H_2} \rangle$, $\langle x_{H_2^+}
\rangle$ and $\langle x_{H^-} \rangle$ as a function of redshift for
the 64L05p2noLW, 64L05p2 and 64L05p3 runs. These runs have the same
parameters ($\epsilon_*=0.2$, $(\epsilon_{UV}/4\pi)=1.6 \times 10^{-5}$, and
\fesc$=1$) except for a different flux in the Lyman-Werner
bands\footnote{ The SED, $g_\nu$, of ionizing radiation is different
  for \pop3 and \popII, but since $\epsilon_{UV}$ is fixed, the \HI
  ionization rate is the same.}. The solid line shows the 64L05p2noLW
run for which the stars do not emit Lyman-Werner photons. The
short-dashed line shows the 64L05p2 run that has a \popII SED, and the
long-dashed line shows the 64L05p3 run that has a \pop3 SED (zero
metallicity). The \popII SED emits about 100 times more photons (per
M$_\odot$ of SF) in the Lyman-Werner bands than the \pop3 SED.
Indeed, in the 64L05p2 run, the dissociating background reduces the
\H2 abundance in the low-density IGM to a value 1/100 of the \H2
abundance in the 64L05p3 run. In the 64L05p2noLW the \H2 is not
destroyed. In Figure~\ref{fig:r2}~(bottom) we show the SFR as a
function of redshift for the same simulations.  Surprisingly, the SFR
does not depend appreciably on the intensity of the dissociating
background. It is evident that the destruction of H$_2$ in the
low-density IGM does not affect the global SFR. The only difference in
SF history is the run with a \popII SED, which has the higher
background in the Lyman-Werner bands. Here, the SFR decreases more
than in the other two runs before a new burst occurs.  However, on
average, also taking into account numerical errors, the SFR is
indistinguishable in these three runs.

\def\capfigd{(Top) Species abundances for the 64L05p2noLW (solid line),
  64L05p2 (short-dashed line), and 64L05p3 (long-dashed line) runs. These
  runs have the same parameters ($\epsilon_*=0.2$, $(\epsilon_{UV}/4\pi)=1.6
  \times 10^{-5}$ and \fesc$=1$) except for a different flux in the
  Lyman-Werner bands. The two top panels show the volume
  (left) and mass (right) weighted \H2 abundance in the box. The
  middle and bottom panels show the analogous quantities for H$_2^+$
  and H$^{-}$ respectively. (Bottom) The comoving SFR for the same
  three runs. Despite the fact that the dissociating background
  flux in these three simulations differs by several orders of
  magnitude, the SFRs are almost identical.}
\placefig{
\begin{figure*}[thp]
\epsscale{0.5}
\plotone{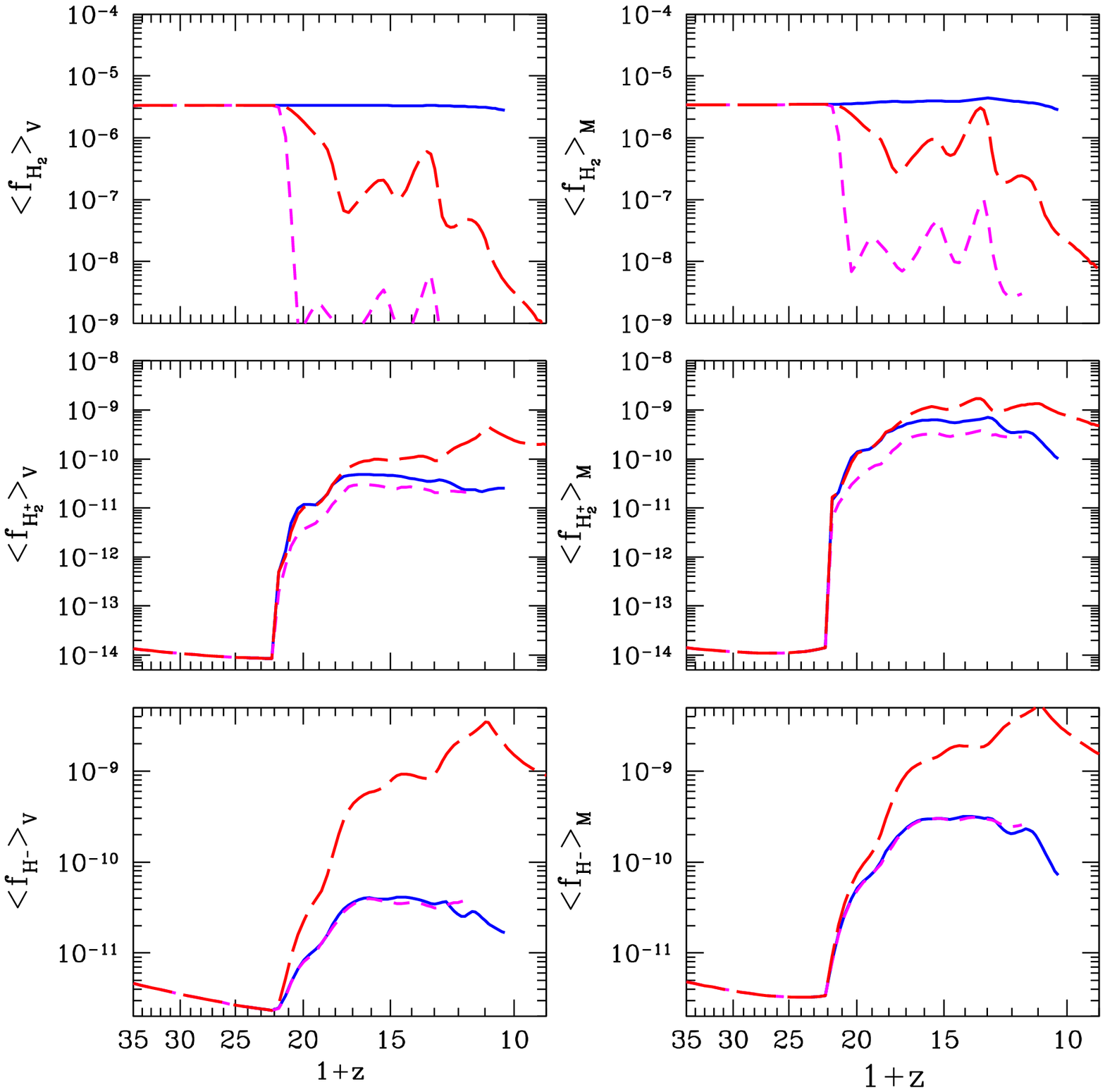}
\plotone{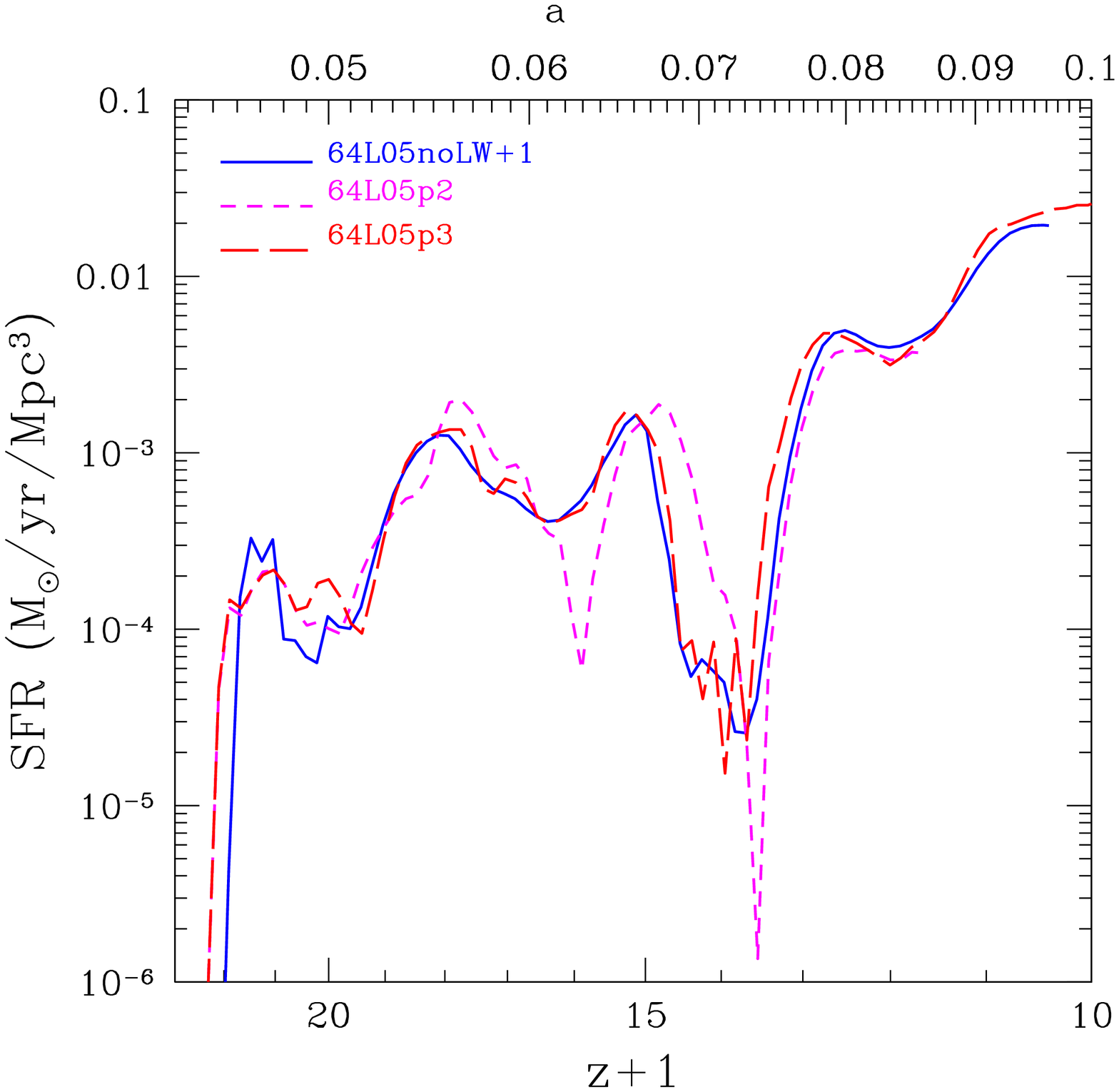}
\epsscale{1.}
\caption{\label{fig:r2}\capfigd}
\end{figure*}
}

When the background in the Lyman-Werner bands (averaged over 1000-1100
\AA) builds up to $J_{LW} \simgt 10^{-25}$ \ju (at $z \sim 20$), it
starts to dissociate the \H2 in the IGM. It is evident
(Fig.~\ref{fig:r2}) that \H2 is continuously destroyed and reformed.
The \H2 abundance peaks are slightly delayed with respect to the SFR
minima.  Since SF is self-regulated, the dissociating background
intensity, after a rapid build-up phase, reaches an almost constant
value.  Depending on the SED of the sources and \fesc, the equilibrium
value of the background can be small (\eg, \pop3 and high \fesc) or
large (\eg, \popII and small \fesc).  Consequently, the \H2 abundance
in the IGM can assume either a large or small quasi-constant value
during the self-regulated SF phase.

In the \pop3 SED run (long-dashed line), the molecular abundance in
the IGM is 100 times higher than in the \popII run, and the
mass-weighted abundance becomes $\langle x_{H_2} \rangle_M \approx 2
\times 10^{-6}$ at $z \sim 12$. This means that the positive feedback
produced essentially the same mass of \H2 destroyed by negative
feedback.

The opacity of the IGM in the Lyman-Werner bands is proportional to
the mean \H2 number density. In Figure~\ref{fig:radLW} we show the
background specific intensity in $10^{-21}$ \ju units, $J_{21}(\nu)$,
as a function of frequency for the 64L05p3. We show $J_{21}(\nu)$ at
redshift $z = 12$ when $\langle x_{H_2} \rangle_M \sim 2 \times
10^{-6}$. In the upper panel, it is interesting to note the importance
of the spectral features caused by \HI, \GI, and \GII Ly$\alpha$
emission lines. The lower panel shows a zoom at the frequencies of the
Lyman-Werner bands; the upper (lower) line shows the intensity of the
background without (with) line radiative transfer. The \H2 and
resonant \HI Lyman series line opacities reduce the background
intensity by about one order of magnitude.

\def\capfige{Background specific intensity in $10^{-21}$ \ju units,
  $J_{21}(\nu)$, as a function of photon energy for the 64L05p3 run
  (\pop3) at $z = 12$. The lower panel shows a zoom of the spectrum in
  the Lyman-Werner bands. In this example, \H2 and H Lyman series line
  opacities reduce the intensity of the dissociating background in the
  Lyman-Werner bands by about one order of magnitude (lower line
  compared to the upper one).}
\placefig{
\begin{figure*}[thp]
\plotone{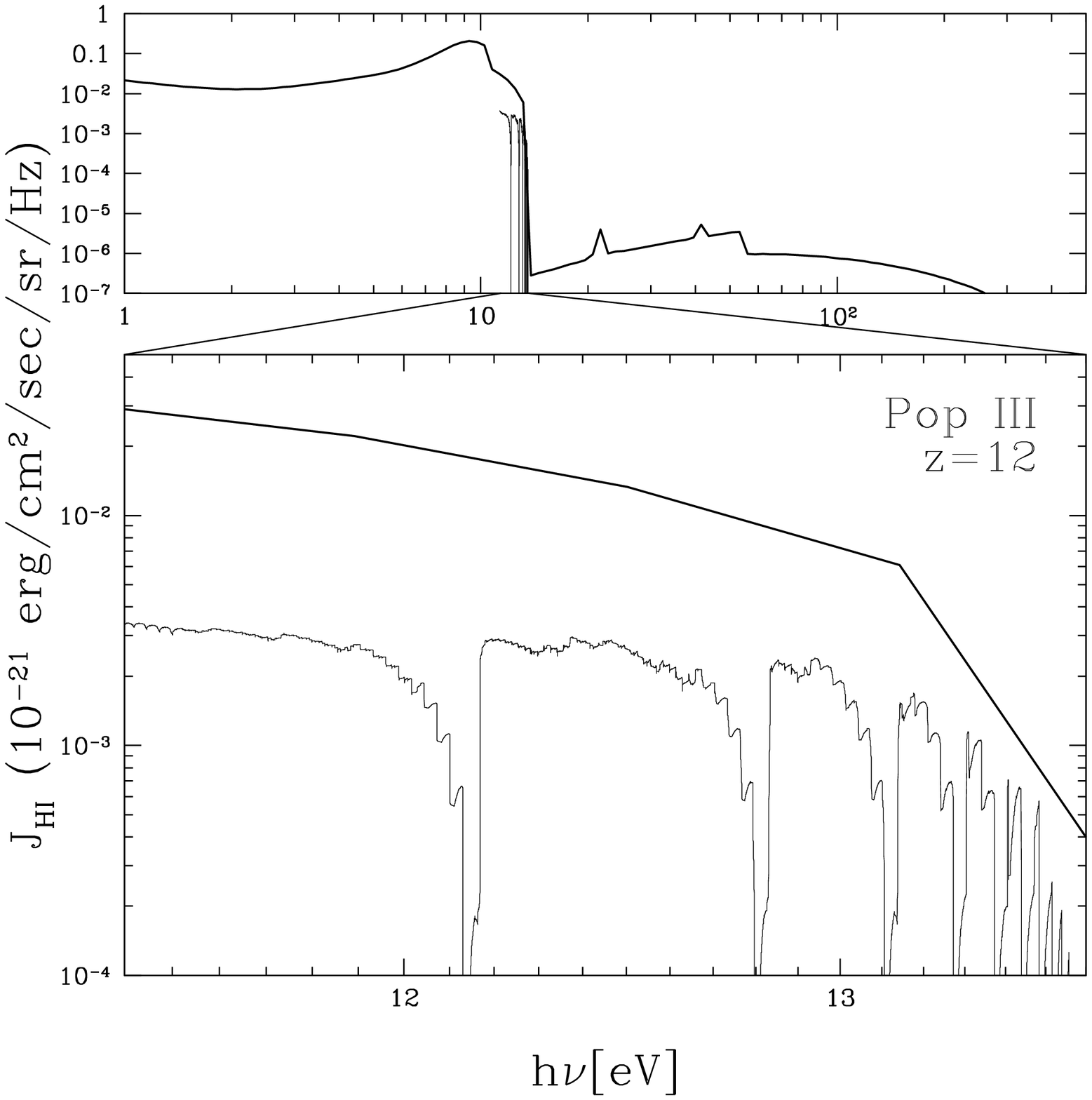}
\epsscale{1.}
\caption{\label{fig:radLW}\capfige}
\end{figure*}
}

In Figure~\ref{fig:slice1} we show a time sequence of two slices
through the most massive object for the 64L05p2 simulation at $z=19.4$
and 18.5. This simulation has $\epsilon_*=0.2$, $(\epsilon_{UV}/4\pi)=1.6
\times 10^{-5}$, \fesc$=1$, and sources with a \popII SED. Each one of
the two panels shows $\log x_{\HI}$, $\log x_{H_2}$, $\log
(1+\delta)$ and $\log T$, where $\delta=(\rho-\rho_0)/\rho_0$ is the
baryon overdensity with respect to the mean IGM density $\rho_0$. At
$z=19.4$, the \H2 has its relic abundance everywhere in the IGM except
in the dissociation spheres around the first objects, where it is
destroyed. At $z=18.5$, the dissociation spheres are still visible,
but the UV background starts to dissociate \H2 everywhere in the IGM
except in the filaments.  Finally, at $z \simlt 18.5$, before the
dissociation spheres overlap, the background has destroyed all
the relic \H2. The \H2 is still present in the filaments where the gas
is partially ionized by stars, and positive feedback dominates.  In the
analogous simulation, where the sources have a \pop3 SED, the
dissociation spheres never appear around the source. The dissociating
background destroys the \H2 in the IGM before the dissociation spheres
grow larger than the PFRs.  Finally, if \fesc$<1$ and we use a \popII
SED for the sources, the dissociation spheres around the sources
almost overlap before the background dominates the \H2 dissociation
rate. If the dissociating radiation emitted by each source, $S_{LW}$,
is large, the dissociating background intensity rises quickly above
$J_{LW} \approx 10^{-25}$ \ju and the dissociation of \H2 in the IGM
happens abruptly. In this case, the dissociation spheres
grow fast enough\footnote{In \cite{RicottiGS:01} we provide an analytic
expression for the comoving radius, $R_D$, of the dissociation sphere
produced by a source that turns on at $z=z_i$ as a function of time:
$R_D \propto (z_i+1)(S_{LW}t)^{1/2}$.} to cover a large volume of the
IGM before the contribution of distance sources to the dissociating
background builds up substantially. In contrast, if the dissociating
radiation emitted by the sources is small, the dissociation spheres
grow slowly, while the the additive contribution of distant sources
builds up the intensity of the dissociating background more quickly.
In this case, the dissociation spheres remain smaller than the PFRs
(and therefore invisible) until the dissociating background has
destroyed all the \H2 in the IGM.
  
\def\capfigf{Slices through the most massive
  object in the simulation of the 64L05p2 box at $z=19.4$ and
  18.5. The box size is $L_{box}=0.5$ comoving \Mpc.
  Each one of the two panels shows $\log x_{\HI}$ (top-left), $\log
  x_{H_2}$ (top-right), $\log (1+\delta)$ (bottom-left) and $\log T$
  (bottom-right).}
\placefig{
\begin{figure}[thp]
\epsscale{1.}
\plotone{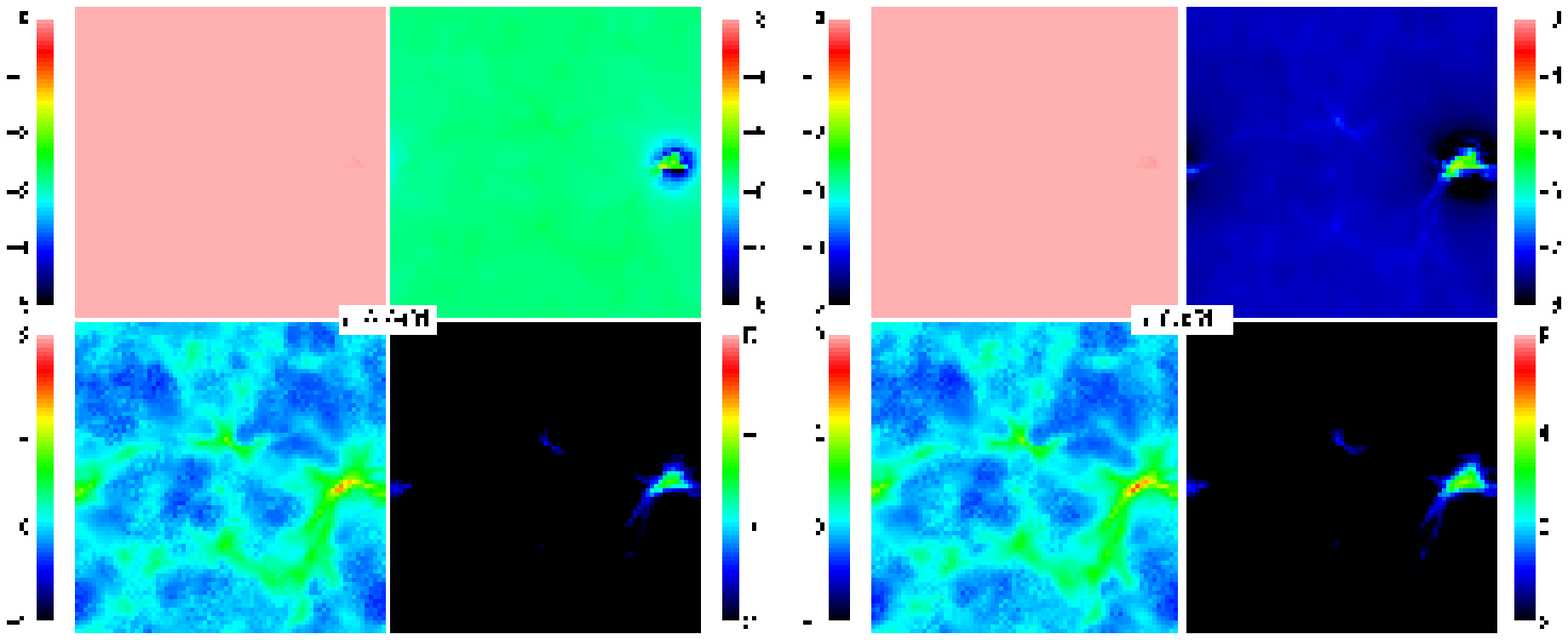}
\caption{\label{fig:slice1}\capfigf}
\end{figure}
}

Figure~\ref{fig:slice2} is analogous to Figure~\ref{fig:slice1},
except that we now show a zoomed region ($0.125 \times 0.125$ \Mpc)
around the most massive object in the 64L05p3 box.  This simulation
has $L_{box}=0.5$ \Mpc, $\epsilon_*=0.2$, $(\epsilon_{UV}/4\pi)=1.6 \times
10^{-5}$, \fesc$=1$, and sources with a \pop3 SED.  In this sequence of
four slices (at $z= 17.3, 12.2, 11.3$, and 10.2) we recognize the two
main processes that create \H2 in the filaments. In the top-left frame
at $z=17.3$ we can recognize a PFR as a shell of \H2 surrounding the
\HII region that is barely intersected by the slice. In the
bottom-left frame ($z=11.3$) two \HII regions are clearly visible.
Inside the \HII regions, the \H2 is destroyed. In the bottom-right
frame ($z=10.2$) the \HII regions are recombining (demonstrating that
the SF is bursting) and new \H2 is being reformed inside the relic \HII
regions. A more fine inspection\footnote{Movies of 2D slices and 3D
  rendering of the simulations are publicly available on the web at
  the URL: http://casa.colorado.edu/$\sim$ricotti/MOVIES.html} of the
time evolution of this slice shows that at least five \HII regions
form and recombine between $z = 20$ and 10 in this small region of the
simulation.

\def\capfigg{ Same as in Figure~\ref{fig:slice1} except for a zoomed
  region of $0.125^2$~$h^{-2}$~Mpc$^2$ around the most massive object
  in the 64L05p3 simulation. In this time sequence of images (top:
  $z=17.3, 12.2$ from left to right; bottom: $z= 11.3, 10.2$ from left
  to right) we recognize the two main processes that create \H2 in the
  filaments: PFRs in front of \HII regions and the reformation of \H2
  inside relic \HII regions. The bursting mode of the SF is evident
  from the continuous formation and recombination of the \HII regions
  in the time sequence of the slices.}  
\placefig{
\begin{figure*}[thp]
\epsscale{1.}
\plotone{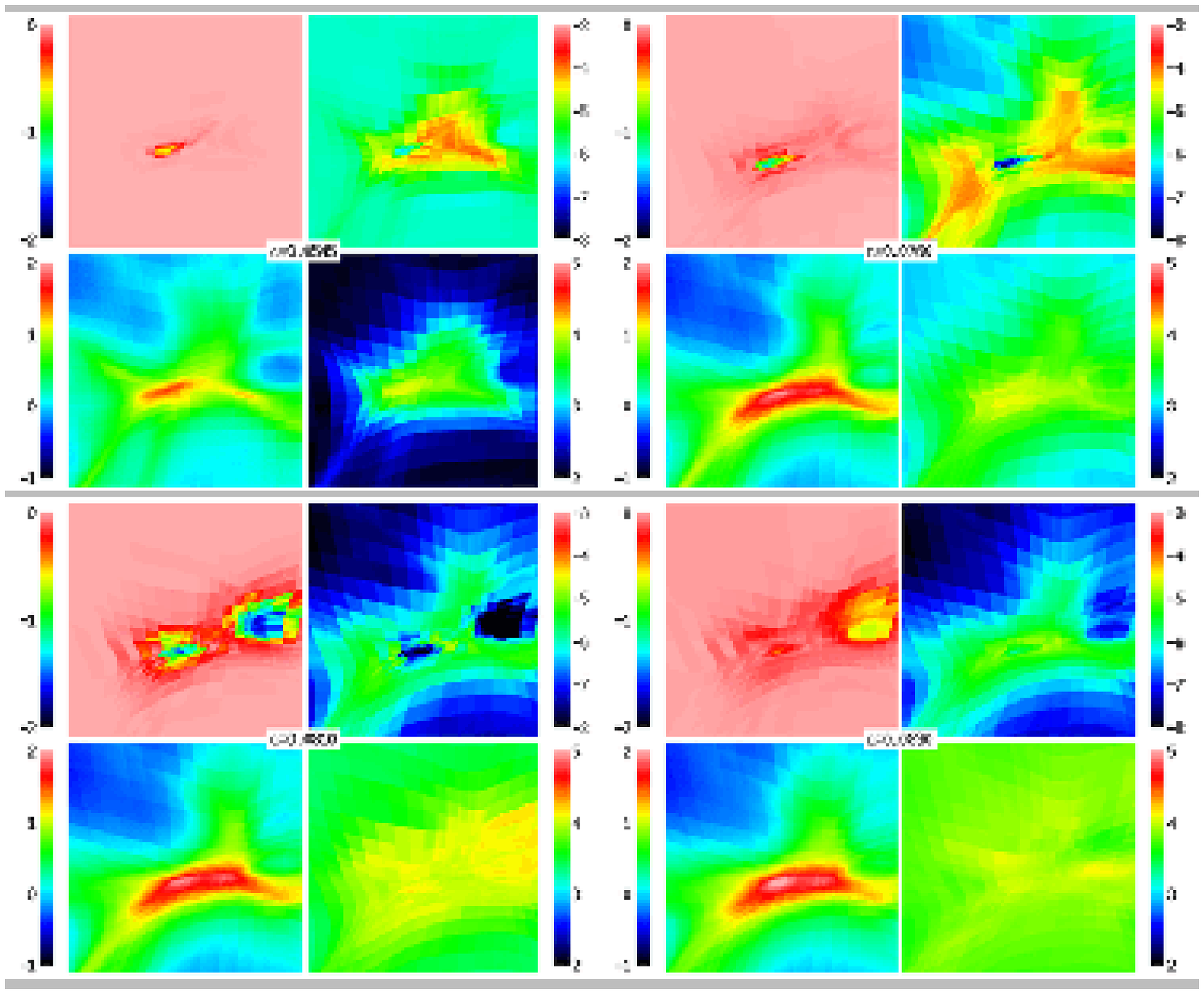}
\caption{\label{fig:slice2}\capfigg}
\end{figure*}
}

If $\epsilon_*(\epsilon_{UV}/4\pi)$\fesc$ \ll 3.2 \times 10^{-6}$ the SF mode
is less violently bursting, and the radiative feedback does not
suppress the SFR with respect to the case without radiative transfer. In
this case, the positive feedback should be dominated by PFRs
preceding the ionization fronts rather than by PFRs inside relic \HII
regions. From the inspection of some 3D movies showing the formation
of \HII regions, it appears that galaxy formation could be triggered by
the presence of neighboring galaxies, in a chain-like process. It is
difficult to prove quantitatively this effect, given that galaxies tend
to be clustered and, in the early universe, galaxies of very different
mass can virialize at the same redshift.

\subsection{Self-Regulated Star Formation History\label{ssec:sim}}

In Figure~\ref{fig:r3}~(left) we compare the SFR of three simulations
with $L_{box}=0.5$ \Mpc, $(\epsilon_{UV}/4\pi)=1.6 \times 10^{-5}$,
\fesc$=1$, and a \pop3 SED (64L05p3, 64L05p3b, 64L05p3c), when we
reduce $\epsilon_*=0.2$ by a factor of 10 and 100. It appears that the
SFR is fairly insensitive to the value of $\epsilon_*$ if we consider
a realistic range of values $0.2 < \epsilon_* <0.02$ . As $\epsilon_*$
is reduced from 0.2 to 0.02, the oscillations of the global SFR with
redshift become more smooth, but its redshift-averaged value gets only
a factor of two smaller. When we reduce $\epsilon_*$ from 0.02 to
0.002, the global SFR gets a factor of 5 smaller. It is important to
note that, when the SFR is dominated by \pII objects, the global SFR
is proportional to $\epsilon_*$. This result is based on the
comparison of several simulations that differ only on the value of
$\epsilon_*$, some presented here and some presented in
\cite{Gnedin:00}.  

The \HI ionizing background, $J_{\HI}$, has the same behavior as the
SFR (\ie, is insensitive to $\epsilon_*$). In
Figure~\ref{fig:r3}~(right) we show the comoving mean free
path\footnote{By definition, $\lambda_{com}=[c/\overline k_\HI(1+z)]$
where $\overline k_\HI$ is the mean absorption coefficient weighted by
the photoionization rate. Neglecting the terms on the left hand side
of equation~(4) in paper~I, we have $\overline k_\HI=\overline S_\HI
/\overline J_\HI$. Therefore we can derive $\lambda_{com}$ from the
emissivity $\overline S_\HI$ (that is proportional to the SFR, \fesc
and $\epsilon_{UV}$) and the ionizing background intensity
$\overline J_\HI$.} of \HI ionizing photons $\lambda_{com}$ for
the same simulations shown in Figure~\ref{fig:r3}~(left). It is
striking that, even if we change $\epsilon_*$ by two orders of
magnitude, the mean free path of ionizing photons oscillates around a
constant value, $\lambda_{com}^{cr}$, shown in the figure by the solid
line given by
\begin{equation}
\lambda_{com}^{cr} = 0.55~h^{-1}~{\rm kpc}\left({20 \over 1+z}\right)^2.
\label{eq:lcri}  
\end{equation}

\def\capfigh{ Global SFR (left) and comoving mean free path of \HI
  ionizing photons, $\lambda_{com}$ (right) as a function of redshift
  for the 64L05p3, 64L05p3b and 64L05p3c runs. The solid,
  short-dashed, and long-dashed lines have $\epsilon_*=0.2, 0.02$, and
  0.002, respectively. The three simulations have $L_{box}=0.5$ \Mpc,
  $(\epsilon_{UV}/4\pi)=1.6 \times 10^{-5}$, \fesc$=1$, and a \pop3 SED.
  The global SFR, $J_{\HI}$, and $\lambda_{com}$, are almost insensitive to
  $\epsilon_*$ for a reasonable range of SFEs ($0.2 < \epsilon_* < 0.02$).}
\placefig{
\begin{figure*}[thp]
\plottwo{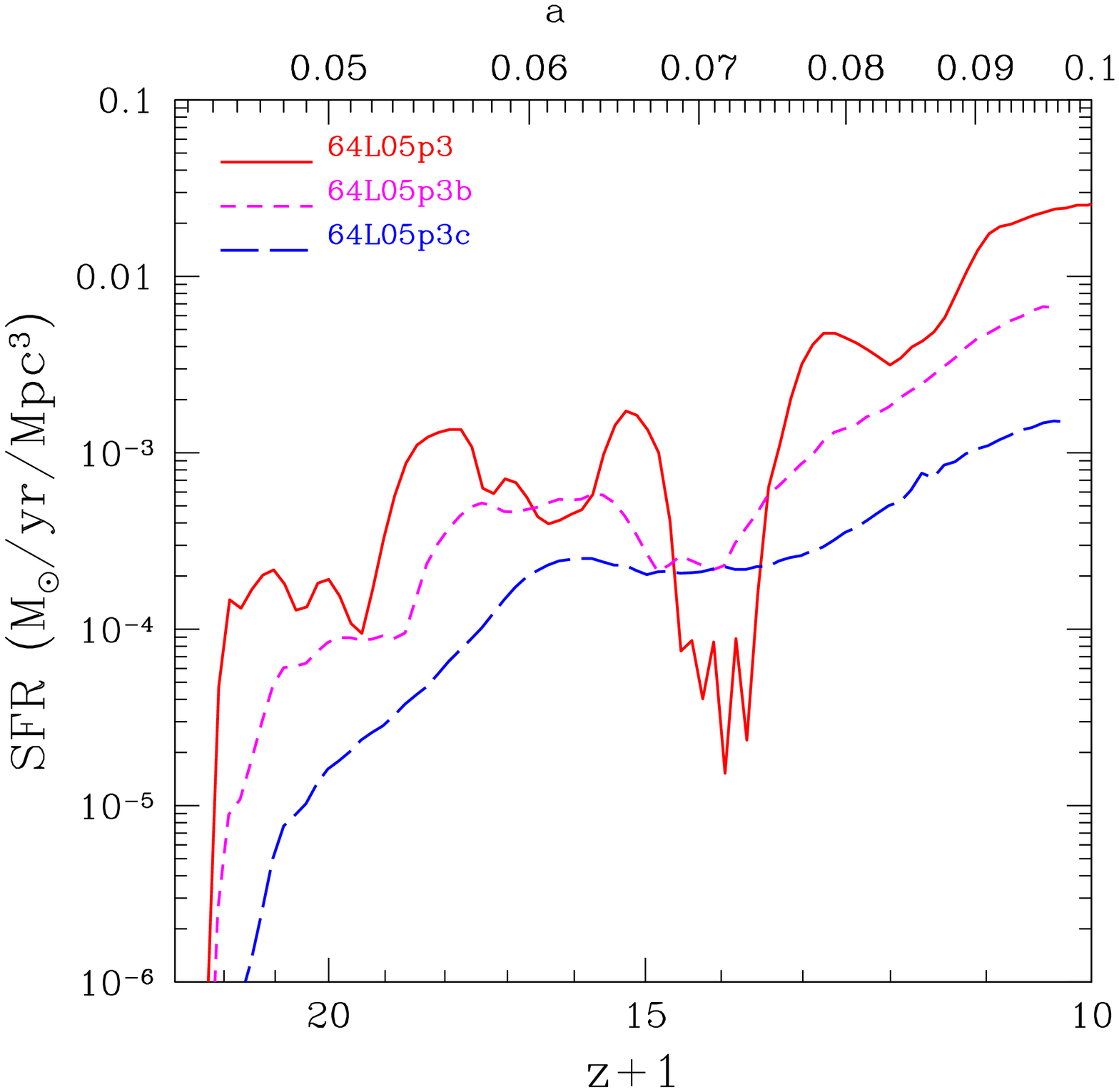}{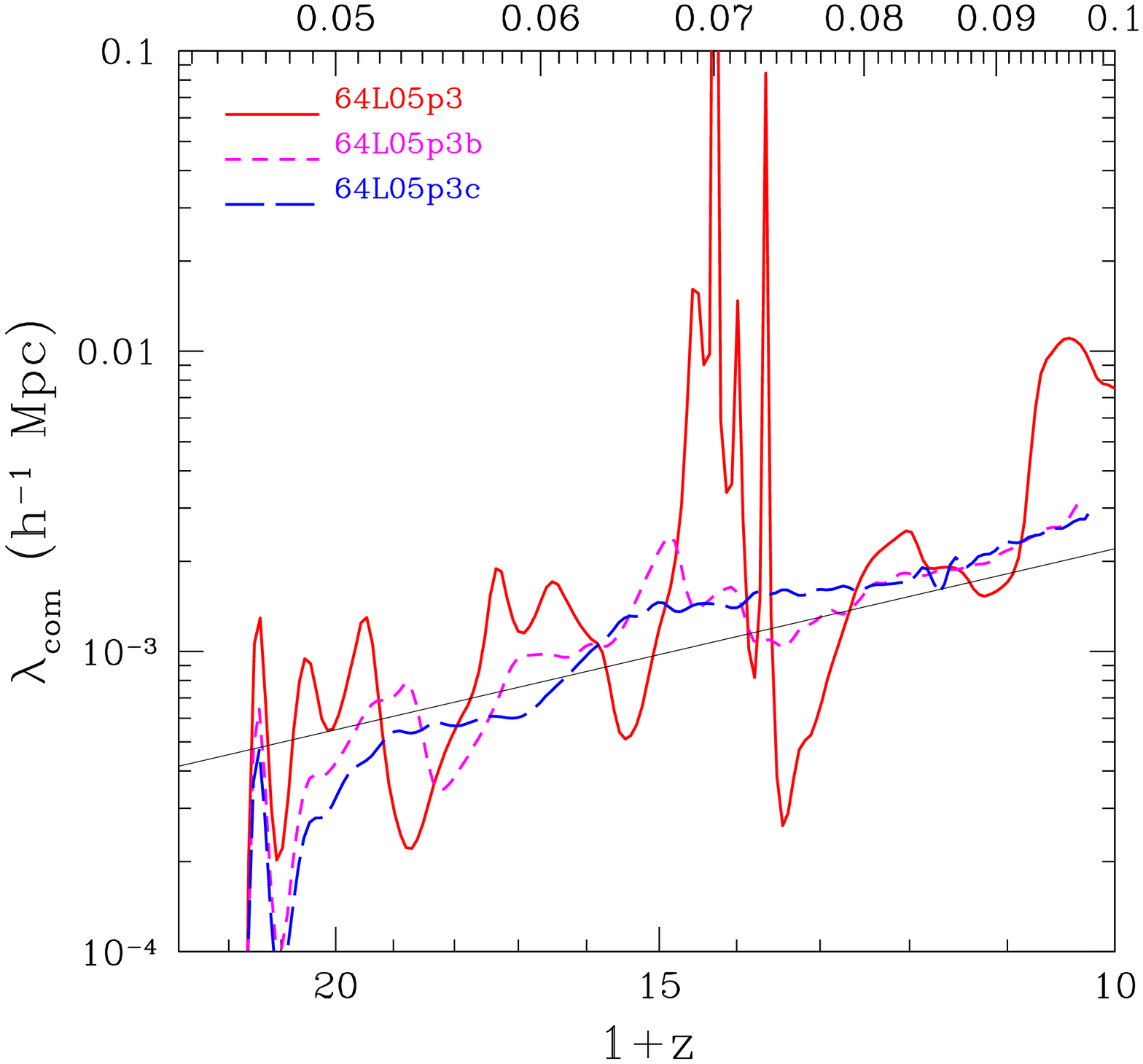}
\caption{\label{fig:r3}\capfigh}
\end{figure*}
}

Figure~\ref{fig:r4} is analogous to Figure~\ref{fig:r3}, but compares
simulations varying the value of $\epsilon_{UV}$. The first three runs
of the list shown in the figure (thick lines) have the same \fesc$=1$
and $\epsilon_*=0.2$, but $\epsilon_{UV}$ is reduced by factors of 10
and 100; the solid, short-dashed and long-dashed lines have
$(\epsilon_{UV}/4\pi)=1.6 \times 10^{-7}, 1.6 \times 10^{-6}$ and $1.6 \times
10^{-5}$, respectively. The last three runs of the list (thin lines),
have $\epsilon_*=0.02$, and varying $\epsilon_{UV}$; the solid,
short-dashed and long-dashed lines have $(\epsilon_{UV}/4\pi)=1.6 \times
10^{-6}, 1.6 \times 10^{-5}$ and $1.6 \times 10^{-4}$,
respectively. Figure~\ref{fig:r4}~(left) shows that the SFR is
approximatively inversely proportional to $\epsilon_{UV}$ and
insensitive to $\epsilon_*$. The inverse proportionality relation is
evident when the SF is smooth (compare the thick and thin solid
lines). When the SF is bursting the comparison of different
simulations is more difficult but the inverse relation appears to hold
at least in a limited range of the parameter
space. Figure~\ref{fig:r4}~(right) shows that $\lambda_{com}$ is
constrained to not exceed a critical value, $\lambda_{com}^{cr}$,
shown by the solid line of equation~(\ref{eq:lcri}). Analogous to
$\lambda_{com}$, $J_{\HI}$ is insensitive to the choice of the free
parameters of the simulations.  The number of ionizing photons that
escape in the IGM is propotional to the parameter combination
$\epsilon_{UV}$\fesc. Therefore changing $\epsilon_{UV}$ is the same
as changing \fesc. The only difference is that the SED has more
dissociating photons if we reduce \fesc insted of $\epsilon_{UV}$. In
\S~\ref{subsec:fesc} we show that, unless \fesc is very small, the
dissociating radiation does not affect the SFR.  In this regime is the
value of the parameter combination $\epsilon_{UV}$\fesc that regulates
the SFR.

If the product $\epsilon_* \epsilon_{UV}$\fesc is large,
$\lambda_{com}$ has large oscillations around the critical value, and
when the product is small, the oscillations are small. Large
oscillations of $\lambda_{com}$ are associated with a strong bursting
SF mode. The feedback works in such a way that when $\lambda_{com}$
exceeds the critical value the SF is suppressed, and consequently
$\lambda_{com}$ becomes smaller than the critical value.  It is easy
to show that $\lambda_{com}$ is related to the the \HII region radii.
Therefore, the mechanism that self-regulates the SF in \p3 objects is
related to the size of \HII regions, rather than to the intensity of
the dissociating background as previously thought.

Following \cite*{Gnedin:00}, we can express $\lambda_{com}$ as a
function of the mean radius, $R_\HII$, of the \HII regions well before
the overlap phase. By definition, $\lambda_{com}=[c/\overline k(1+z)]$
where $\overline k$ is the mean absorption coefficient weighted by the
photoionization rate $\Gamma$. The mean optical depth for \HI ionizing
radiation inside an \HII region is $\langle \tau \rangle=
R_{\HII}/\lambda_{com} = \langle \sigma_\HI N_{\HI} \rangle$, where
$\sigma_\HI$ is the \HI ionization cross section and $N_{\HI}$ is the
\HI column density of the halo. The \HI column density of a virialized
halo is
\[
N_{\HI} = x_\HI n_{vir} R_{vir} \sim 10^{19}~{\rm
  cm}^{-2}\left({1+z \over 20}\right)^2.
\]
Here $n_{vir}$ is the virial density, $R_{vir}$ is the virial radius
of a $M_{DM} \sim 10^7$ M$_\odot$ halo,
and $x_\HI \sim 0.05$ (derived from the simulations) is the hydrogen neutral
fraction. Assuming a frequency-averaged value of the \HI
photoionization cross section
$\langle \sigma_\HI \rangle \sim 10^{-18}$ cm$^2$, we have:
\begin{equation} 
\lambda_{com} = {R^{com}_\HII \over \langle \tau \rangle} =
{R^{com}_\HII \over \langle \sigma_\HI N_\HI \rangle} \approx 0.1 R^{com}_\HII[h^{-1}~{\rm kpc}] \left({20 \over 1+z}\right)^2,
\label{eq:lcom}
\end{equation}
where $R_\HII^{com}$ [\kpc] is the mean free path of ionizing
photons.  Comparing equation~(\ref{eq:lcom}) with
equation~(\ref{eq:lcri}), we find that the mean radius of the \HII
regions produced by \p3 objects is $R_\HII^{com} \approx 5$ \kpc,
about the size of the dense filaments and the virial radii of the
halos.

Qualitatively, this result is already evident in
Figure~\ref{fig:tile}, which shows that the \HII regions remain
confined inside the filaments. The PFRs produced ahead of ionization
fronts and the relic \HII regions continuously reform \H2 inside the
filaments. The \H2 abundance remains high in the filaments, even when
the dissociating background intensity is sufficiently strong to
dissociate all the \H2 in the lower-density IGM.

\def\capfigi{Same as in Figure~\ref{fig:r3} for the 64L05p2noLW-2,
  64L05p2-1, and 64L05p2 runs (thick solid, short-dashed, and
  long-dashed lines) with $\epsilon_*=0.2$, \fesc$=1$, and
  $(\epsilon_{UV}/4\pi)=1.6 \times 10^{-7}, 1.6 \times 10^{-6}, 1.6 \times
  10^{-5}$. The thin solid, short-dashed, and long-dashed lines, show
  the 64L05p3b-1, 64L05p3b, and 64L05p2noLW+1 runs that have
  $\epsilon_*=0.02$, \fesc$=1$, and $(\epsilon_{UV}/4\pi)=1.6 \times 10^{-6},
  1.6 \times 10^{-5}, 1.6 \times 10^{-4}$. The three curves in each
  set of simulations have values of $\epsilon_{UV}$ increased by one
  order of magnitude. It appears that the SFR is about inversely
  proportional to $\epsilon_{UV}$ and independent of
  $\epsilon_*$. The comoving mean free path of \HI ionizing photons,
  $\lambda_{com}$, and $J_{\HI}$, instead, remain constant.}
\placefig{
\begin{figure*}[thp]
\plottwo{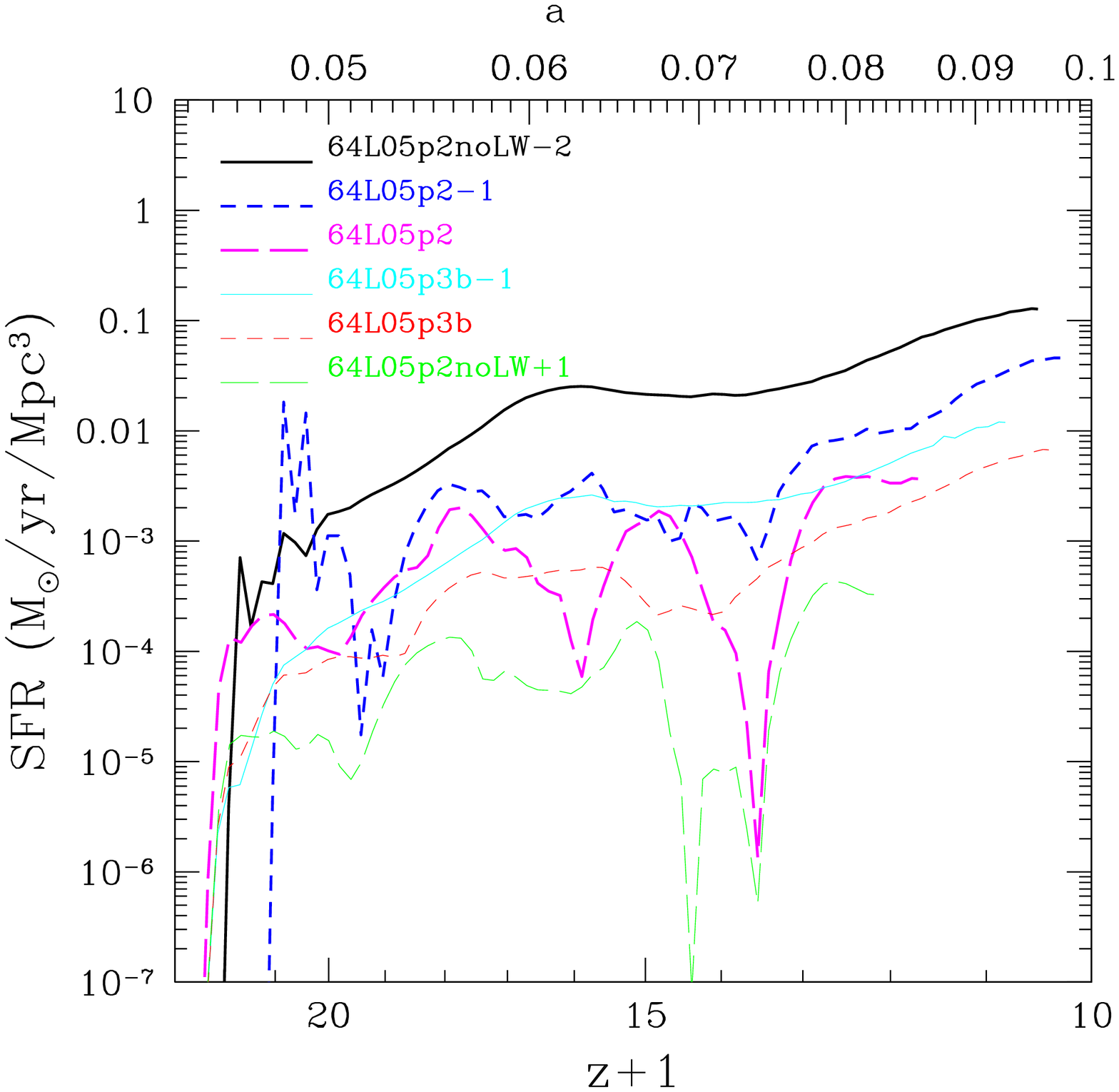}{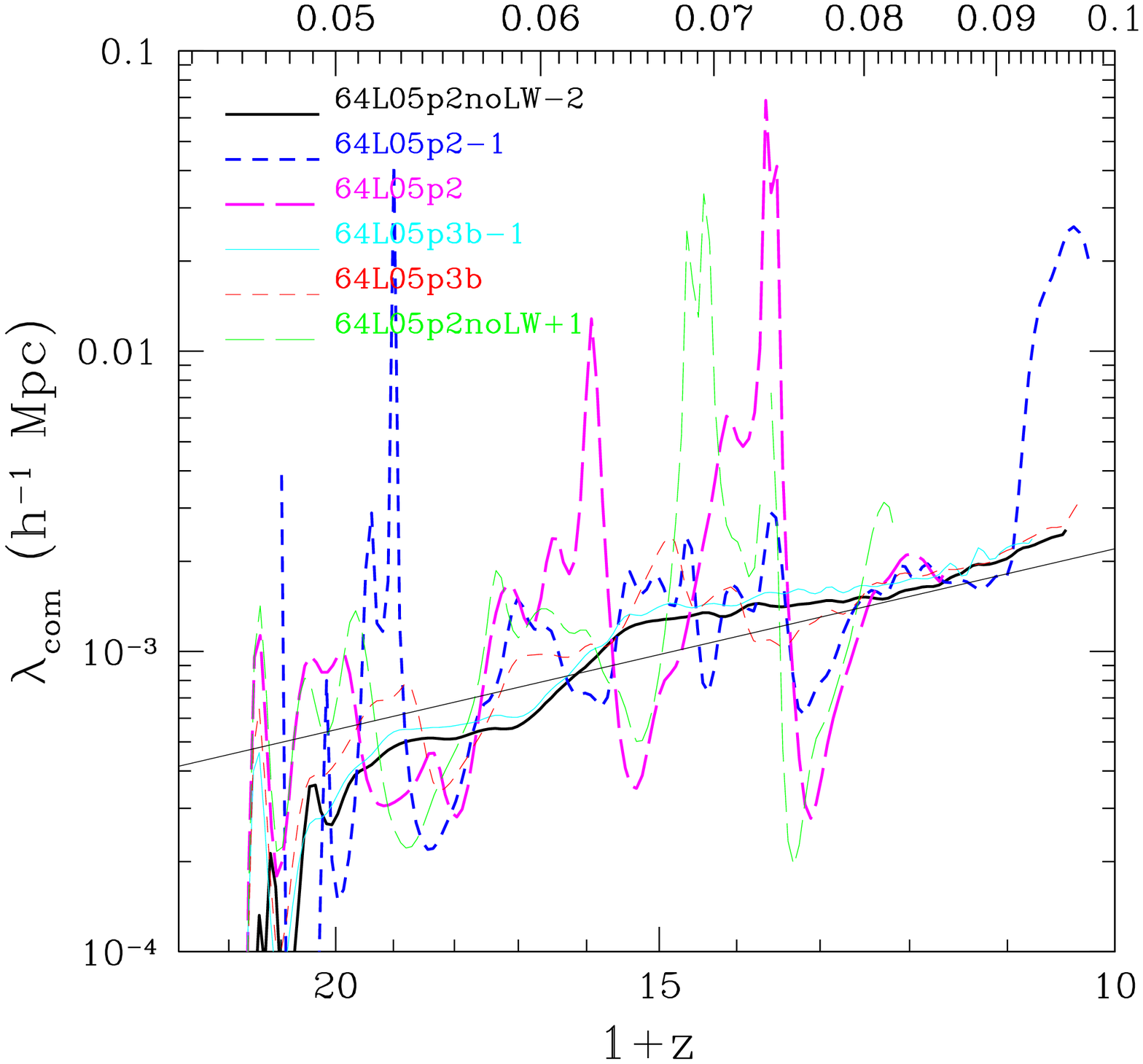}
\caption{\label{fig:r4}\capfigi}
\end{figure*}
}

\subsubsection{Star Formation History: a Function of \fesc\label{subsec:fesc}}

We have seen in the previous section that the SFR is fairly
insensitive to $\epsilon_*$ over the range $0.02 < \epsilon_* < 0.2$
and inversely proportional to $\epsilon_{UV}$\fesc.  The free
parameter $\epsilon_{UV}$ depends mainly on the stellar IMF and
slightly on the stellar metallicity. Assuming a Salpeter IMF we find
$(\epsilon_{UV}/4\pi) = 1.1 \times 10^{-5}$ for a \popII SED and
$(\epsilon_{UV}/4\pi) = 2.5 \times 10^{-5}$ for a \pop3 SED.  Some
theoretical arguments suggest that, at high redshift, the IMF could be
flatter than a Salpeter IMF. If this is the case, we would have
$(\epsilon_{UV}/4\pi) \gg 10^{-5}$. The possibility of having a steeper IMF,
and therefore $(\epsilon_{UV}/4\pi) \ll 10^{-5}$, is not supported by any
theoretical work or observation.

The value of \fesc is unknown, and has been an argument of debate form
many years. In literature, values of \fesc between 0.5 to zero have
been proposed.  Theoretical work on \fesc for \p3 objects at
high-redshift \citep{RicottiS:00, Wood:00} finds that \fesc should be
very small ($10^{-2}<$\fesc$<10^{-5}$) and decreasing with increasing
halo masses.  Observations of \fesc in nearby starburst galaxies find
values of \fesc$\simlt 10 \%$ \citep*{Leitherer:95, Hurwitz:97,
  Heckman:01, Deharveng:01}, in agreement with theoretical studies
\citep*{Dove:00}.  Numerical simulations of the reionization of the
IGM usually adopt a value of \fesc$ \sim 1$ in order to reionize the
universe before redshift $z \sim 6$. A recent study on Lyman break
galaxies \citep*{Steidel:01} agrees with numerical simulations in
finding \fesc$\sim 0.5-1$. But there is no observational constraint on
the value of \fesc for \p3 objects.  Reasonably, \p3 objects should
have smaller \fesc than \pII objects (normal galaxies) since their
halos are not collisionally ionized. On the other hand, if a
substantial fraction of the ISM is photoevaporated or blown away by
stellar winds and SNe, \fesc could be larger. We note that \fesc in
this paper is defined as the escape fraction of ionizing photons from
the resolution element; therefore it is resolution dependent and
generally larger than \fesc from the galactic halos.

In the previous section we showed that, if $(\epsilon_{UV}/4\pi)$\fesc$
\simgt 10^{-5}$ and $\epsilon_* \simgt 0.02$, the SFR is suppressed by
the feedback of ionizing radiation (but not from the dissociating
background).  Since it is unlikely that $(\epsilon_{UV}/4\pi)<10^{-5}$, the
SFR can be increased if \fesc$<1$.  As we decrease \fesc, the SFR
increases almost linearly up to a maximum value determined by the SFR
without any feedback. This value is proportional to $\epsilon_*$ and
to the mass resolution of the simulation. In higher mass resolution
simulations, the number of \p3 objects that form stars is larger since
we resolve many more small-mass objects. We believe that our higher
resolution simulation, with $L_{box}=1$ \Mpc, $256^3$ cells, and mass
resolution $M_{DM} =4.93\times 10^3$ \Ms, is close to fully resolving
SF for the case without radiative feedback (we need to resolve each
halo with about 100 DM particles).

Clearly, if \fesc$ = 0$, there should be no positive feedback, but only
the negative feedback of the dissociating background, which should
determine the SFR.  Indeed, if we decrease the value of \fesc below a
critical value, the SFR, after reaching the maximum, starts
decreasing.  This effect is shown in Figure~\ref{fig:fesc}~(left),
where the lines show simulations with $\epsilon_*=0.2$, \popII SED and
\fesc$=1,0.1,0.01, 10^{-3},10^{-5}$ (64L05p2, 64L05p2-1f, 64L05p2-2f,
64L05p2-3f and 64L05p2-5f runs respectivelly).  At $z=15.5$, in the
simulation with \fesc$=0.01$, the intensity of the dissociating
background is about $J_{LW} \sim 5 \times 10^{-21}$ \ju.  The value of
$J_{LW}$ is high enough to decrease the SFR with respect to the case
without radiative transfer by about a factor of two. In
Figure~\ref{fig:fesc}~(right), we show the different importance of the
dissociating radiation feedback for a \popII and a \pop3 SED. All the
simulations in Figure~\ref{fig:fesc}~(right) have \fesc$=0.01$ (runs
64L05p2-2f, 64L05p3-2f, 64L05p2-2fa and 64L05p3-2fa in Table~1).

\def\capfigl{Stellar fraction, $f_{star}$, as a function of redshift.
  (Left) The two thin solid lines show $f_{star}$ for a simulation
  without radiative transfer and without either radiative transfer or
  \H2 cooling. The thick lines show simulations with \fesc$=
  1,0.1,0.01, 10^{-3},10^{-5}$ as shown in the label. Positive
  feedback increases the SFR as we decrease \fesc from 1 to about
  $10^{-3}$. If \fesc$\simlt 10^{-3}$ the dissociating background is
  effective in suppressing the SF. (Right) The thin solid line shows
  $f_{star}$ for a simulation without radiative transfer and
  $\epsilon_*=0.2$; the thick solid and dotted lines show simulations
  with \fesc$=0.01$, and $\epsilon_*=0.2$, using a \popII or \pop3 SED,
  respectively. The thin dashed, thick dashed, and thick
  dot-dashed lines are analogous to the thin solid, thick solid, and
  thick dotted lines, but for $\epsilon_*=0.05$.}
\placefig{
\begin{figure*}[thp]
\plottwo{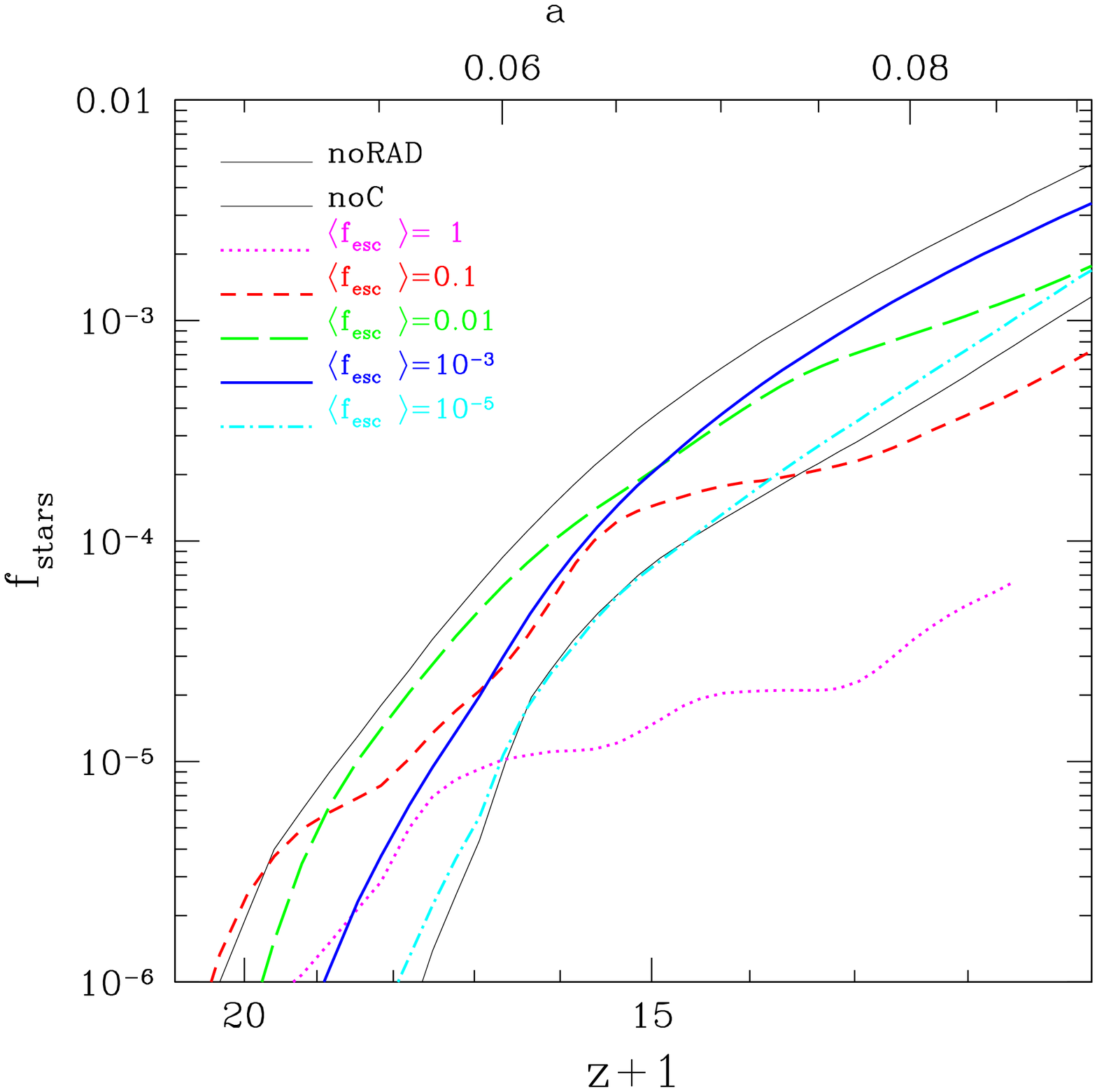}{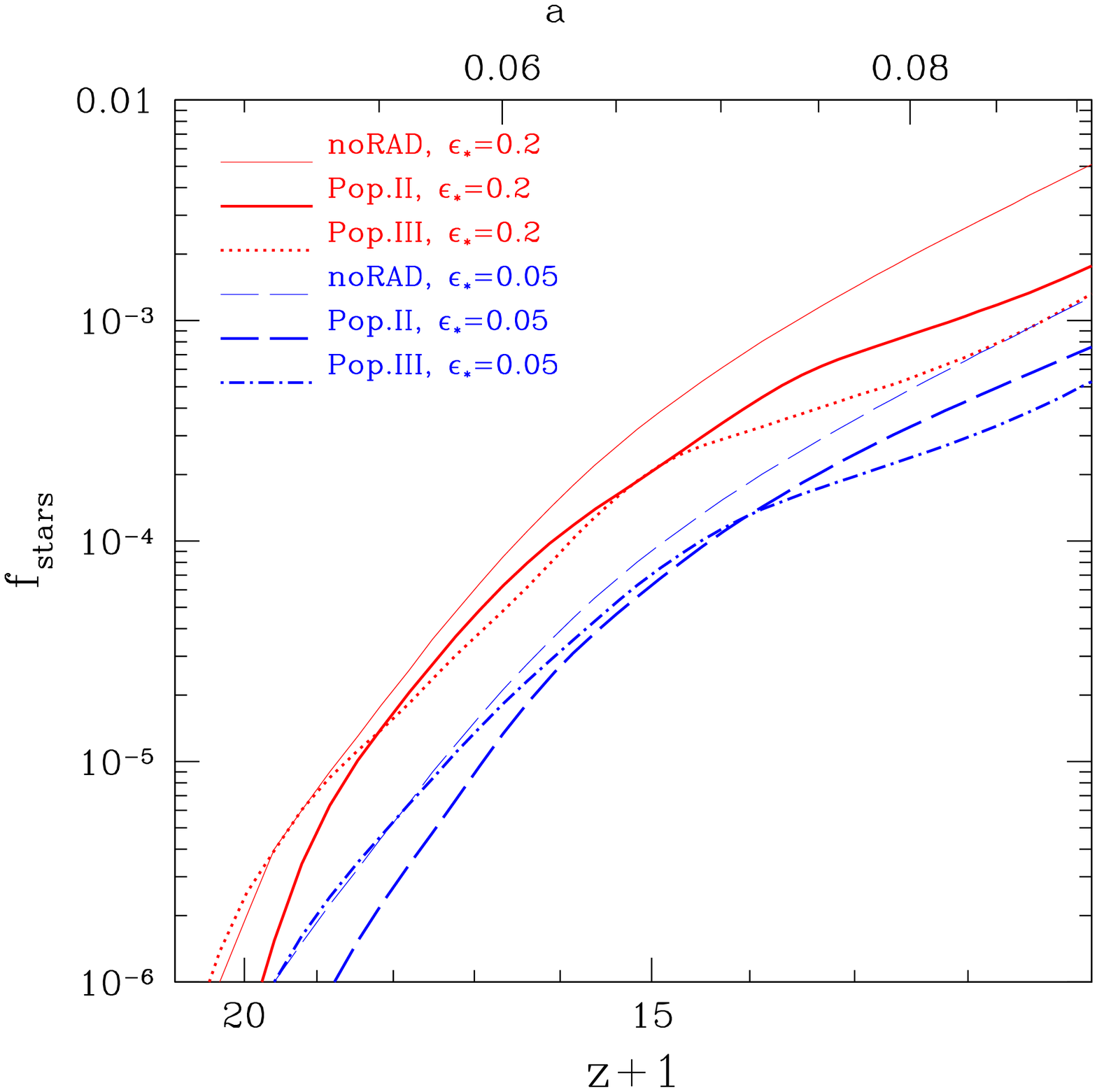}
\caption{\label{fig:fesc}\capfigl}
\end{figure*}
}   

The effect of decreasing excessively the value of \fesc on the SFR is
twofold: (i) it reduces the positive feedback of the EUV radiation;
and (ii) it increases the relative importance of the dissociating
radiation over the ionizing radiation by modifying the SED, $g_\nu$.
The negative feedback of the dissociating background starts to affect
the SFR, depending on magnitude of the jump, $\beta$, at the Lyc
frequency of the SED, and the magnitude of the positive feedback. The
jump $\beta$ is inversely proportional to \fesc and depends on the SED
(intrinsic jump).  For the same \fesc, the \popII SED produces a jump 10
times larger than the \pop3 SED.  We find that, if $0.02 \simlt
\epsilon_* \simlt 0.2$, the critical value for which the negative
feedback of the dissociating background starts to affect the SFR is
\fesc$^{cr} \simlt 10^{-2}$ for the \popII SED, and \fesc$^{cr} \simlt
10^{-3}$ for the \pop3 SED.

In Figure~\ref{fig:fesc3} we show a case where the positive feedback
produces enhanced global SFR with respect to the case without
radiative feedback. The simulation (thick dashed line) has
$\epsilon_*=0.02$, \fesc$=1$, $(\epsilon_{UV}/4\pi)=1.6 \times 10^{-9}$, and
a \pop3 SED (64L05p3b-3 run). The thin solid line shows the analogous
simulation without radiative transfer. The thick dashed line shows a
simulation with the same value of the parameter combination
$\epsilon_{UV}$\fesc as the 64L05p3b-3 run, but with Salpeter IMF
($(\epsilon_{UV}/4\pi)=1.6 \times 10^{-5}$) and \fesc$=10^{-3}$. The SFR in
these two simulations should be identical, since it depends on the
parameter combination $\epsilon_{UV}$\fesc. Instead, the negative
feedback of the dissociating background reduces the SFR by a factor of
three with respect to the 64L05p3b-3 run. The dissociating background
in the simulation with \fesc$=10^{-3}$ is three orders of magnitude
higher ($J_{LW} \sim 0.5 \times 10^{-21}$ \ju at $z=11.5$) than in the
simulation with \fesc$=1$.

\def\capfigm{Stellar fraction, $f_{star}$ as a function of redshift
  for the 64L05noRAD (without radiative transfer, thin solid line),
  64L05p3b-3 (thick solid line), and 64L05p3b-3n (thick dashed line)
  runs. The simulations have $L_{box}=0.5$ \Mpc, $\epsilon_*=0.02$, and
  a \pop3 SED. The 64L05p3b-3 has \fesc$=10^{-3}$ and $(\epsilon_{UV}/4\pi)=1.6
  \times 10^{-6}$. The 64L05p3b-3n run has \fesc$=1$ and
  $(\epsilon_{UV}/4\pi)=1.6 \times 10^{-9}$ (flat IMF). The dissociating
  background is three orders of magnitude higher in the run with
  \fesc$=10^{-3}$.}
\placefig{
\begin{figure*}[thp]
\epsscale{0.8}
\plotone{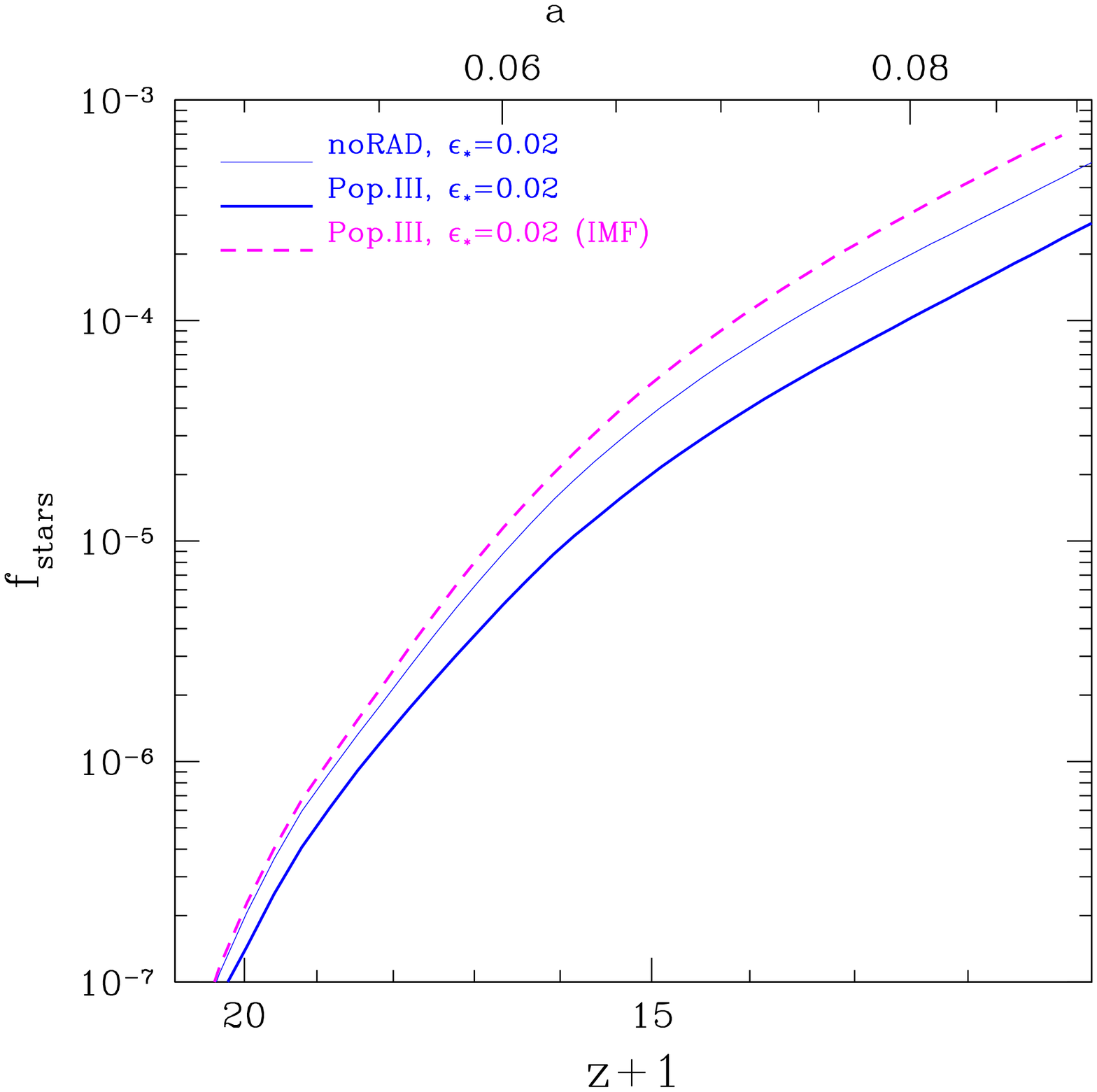}
\epsscale{1.0}
\caption{\label{fig:fesc3}\capfigm}
\end{figure*}
}   

The simulations presented in this section have mass resolution $M_{DM}
= 3.94\times 10^4$ \Ms. The global SFR (without radiative feedback) is
underestimated because we do not resolve the lowest mass objects that
can form stars.  Higher resolution simulations could produce an
enhanced SFR with respect to the case without including any feedback.
Indeed, positive feedback can trigger SF in halos with a mass smaller
than the minimum mass derived in the absence of feedback.  Theoretically,
the lower limit for the DM mass of an object that could form stars is
determined by the filtering\footnote{The filtering mass is an averaged
  Jeans mass that depends on the thermal history of the gas. If the
  temperature remain constant with time, the filtering and Jeans mass
  are identical.} mass that is $M_F \sim 10^5-10^6$ \Ms in the
redshift range $20 < z < 100$. In objects with DM masses smaller than
$M_F$, the baryons cannot virialize.

\subsubsection{How does the Self-Regulation Work?}

Summarizing, the feedback prevents the size of \HII regions from
exceeding $R_{\HII}^{com} \approx 5$ \kpc (about the size of the dense
filaments). Indeed when the \HII regions get bigger than the filaments
molecular hydrogen is destroyed and the SF is suppressed. On the
contrary, when the \HII regions are smaller than the filaments or when
they recombine after a burst of SF, the dense filaments are only
partially ionized and the formation rate of molecular hydrogen is
maximized. Since galaxy formation takes place only in overdense
regions, the SF is self-regulated to maximize \H2 formation in the
filaments.  Thus, the volume filling factor of the \HII regions
remains small. As a result, \p3 objects cannot reionize the universe.
Molecular hydrogen is continuously reformed, in shells preceding the
\HII regions and in shells inside relic \HII regions, the PFRs found
by \cite{RicottiGS:01}. SF is bursting, since it is self-regulated by
the two above-mentioned feedback mechanisms. The SFR does not depend
on $\epsilon_*$ or on the source SED, but only on \fesc and the IMF
through $\epsilon_{UV}$. If $\epsilon_{UV}$\fesc is small, the SFR is
high, and vice versa.  Indeed, if few ionizing photons (per each
baryon converted into stars) escape from the halo, more stars have to
be formed in order to produce \HII regions of the size of the
filaments.  If the product $\epsilon_*\epsilon_{UV}$\fesc is large,
the SF burst is so fast that \HII regions will expand outside the
filaments before recombining.  This produces a temporary halt in the
global SF, which appears as a sequence of strong bursts. It is
fascinating that the SF history of \p3 objects depends primarily on a
single parameter, $\epsilon_{UV}$\fesc. This happens because the
feedback mechanism acts on a cosmological (rather than galactic)
scale. We notice, though, that \fesc should depend slightly on
$\epsilon_*$ \citep{RicottiS:00}.

In Figure~\ref{fig:ske} we show a sketch of the relative importance of
positive and negative feedback from EUV (ionizing) and FUV
(dissociating) radiation on the SFR. The thick and the thin solid
curves show the SFR with and without radiative transfer, respectively,
for a simulation with $\epsilon_*=0.2$, \popII SED ($(\epsilon_{UV}/4\pi)=1.1
\times 10^{-5}$), and \fesc$=0.01$.  Without radiative feedback, the
SFR is proportional to $\epsilon_*$. Therefore, for a simulation with
$\epsilon_*=0.02$, for instance, the SFR shown by the thin solid curve
would be a factor of ten smaller. The solid line shows the region
where the positive feedback from EUV radiation is maximum. The SFR of
this line is inversely proportional to the parameter combination
$\epsilon_{UV}$\fesc.  Above the solid line, we have increasing
negative feedback from EUV radiation (\HII regions become larger than
the filaments), and below the line the positive feedback from EUV
radiation becomes increasingly weak.  If the parameter combination
$\epsilon_*\epsilon_{UV}$\fesc$\simlt 10^{-6}-10^{-7}$, the SFR is not
suppressed by EUV radiative feedback (in this case the solid line lies
above the thin solid curve).  When the flux of ionizing photons in the
filaments is too small, either because \fesc is small or because of a
low SFR, positive feedback effects are weak. The negative feedback of
the dissociating background becomes, therefore, important.  Below the
thin dashed line, the negative feedback from FUV radiation starts to
dominate the positive feedback from EUV radiation.  The region between
the solid and dashed lines is where the positive feedback dominates
over negative feedback, and its size is inversely proportional to the
jump $\beta$ at the Lyc of the source SED. The effect of the FUV
negative feedback is usually to delay SF at high-redshift, when the
SFR, and therefore the ionizing photon flux is lower. At higher
redshift, as the number of \p3 objects increases, the positive
feedback dominates and the SFR becomes self-regulated by the EUV
radiation. In the extreme case of a very small \fesc (\fesc$\simlt
10^{-3}$ for a \popII SED, and \fesc$\simlt 10^{-4}$ for a \pop3 SED),
the negative feedback of the dissociating background can suppress \p3
object formation.

\def\capfign{Sketch showing the relative importance of positive and
  negative feedback from EUV (ionizing) and FUV (dissociating)
  radiation on the SFR. The thick and thin solid lines show the SFR
  with and without radiative feedback, respectively (both for
  $\epsilon_*=0.2$ and \fesc$=0.01$). The shaded strip is where
  positive feedback dominates. Above the strip, the SF is suppressed
  by feedback from ionizing (EUV) radiation, and below from the
  feedback from \H2 dissociating radiation.}
\placefig{
\begin{figure*}[thp]
\epsscale{0.8}
\plotone{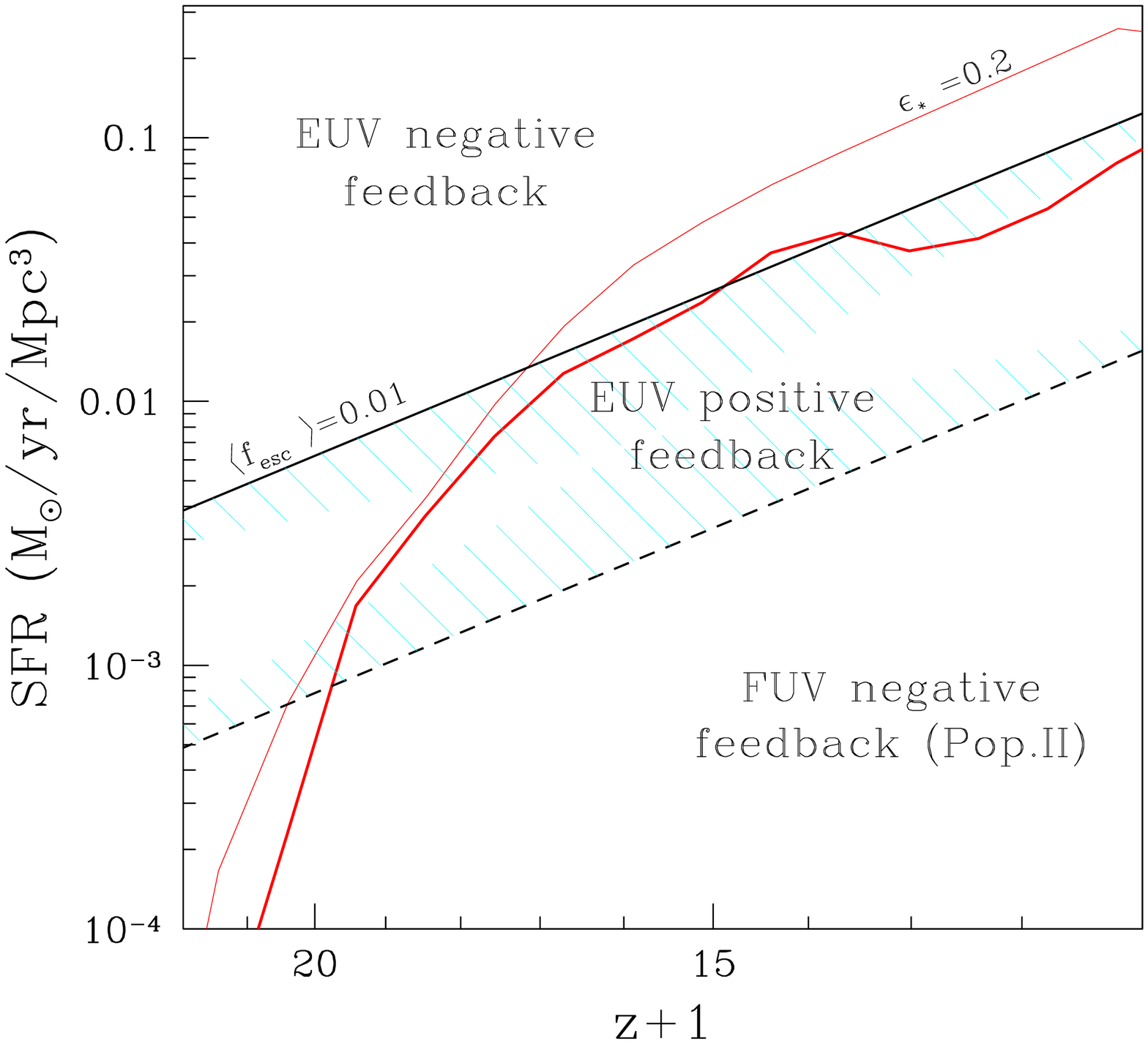}
\epsscale{1.0}
\caption{\label{fig:ske}\capfign}
\end{figure*}
}   

\subsection{Metal Enrichment of the IGM\label{ssec:met}}

Metals are produced by the same massive stars that produce ionizing
radiation. Understanding the metal enrichment of the low density IGM
is a challenging task. Observational constraints from the metallicity
evolution of the Ly$\alpha$ forest allow us to test the models.  In
Figure~\ref{fig:zrend} we show a 3D rendering of the IGM metallicity
(in solar units) for the 64L05p3 run ($L_{box}=0.5$ \Mpc,
$\epsilon_*=0.2$, $(\epsilon_{UV}/4\pi)=1.6 \times 10^{-5}$, \fesc$=1$, and a
\pop3 SED). The opacity and the color coding are proportional to the
logarithm of the metallicity ($Z/Z_\odot$).  The comoving volume
filling factor of the metal enriched gas increases quickly at
high-redshift and slows down as the redshift decreases.
 
\def\capfigo{ 3D rendering of the logarithm of the IGM metallicity
  (in solar units) for the 64L05p3 run. The four cubes show the time
  evolution of the volume filling factor of metal-enriched gas at
  $z=15.7, 13.3, 12.3$, and 10.2.}
\placefig{
\begin{figure*}[thp]
\plotone{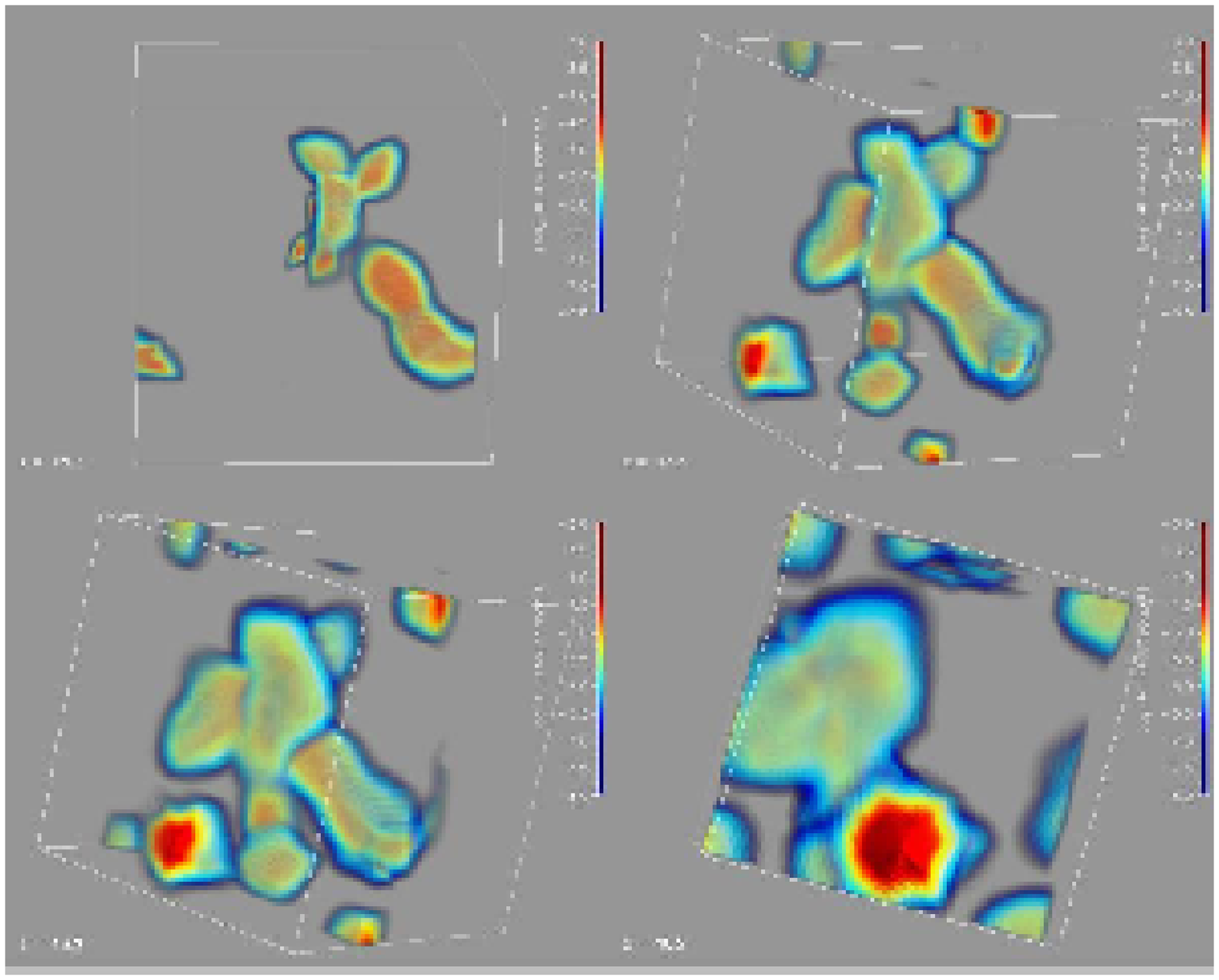}
\caption{\label{fig:zrend}\capfigo}
\end{figure*}
}

In Figure~\ref{fig:zt}~(left) we show the ratio $\langle Z/Z_\odot
\rangle_V/\langle Z/Z_\odot \rangle_M$ of the volume- to mass-weighted
mean metallicities as a function of redshift. This ratio is
proportional to the filling factor of the metal enriched gas. The
solid line shows the 64L05noRAD run ($L_{box}=0.5$ \Mpc,
$\epsilon_*=0.2$, $(\epsilon_{UV}/4\pi)=1.6 \times 10^{-5}$, without
radiative transfer). The dotted, dashed, and long-dashed lines show
simulations with \fesc$=1$ and an increasing value of the parameter
combination $\epsilon_*(\epsilon_{UV}/4\pi)$\fesc$=3.2 \times 10^{-8}, 3.2
\times 10^{-7}$, and $3.2 \times 10^{-6}$. We remind the reader that,
if the value of this parameter combination is high, the global SF,
$J_{\HI}$, and $\lambda_{com}$, are strongly oscillating (bursty SF).
The filling factor of metal-enriched gas is larger in the simulations
with strongly bursting SF.  This suggests that photo-evaporation of
high-redshift \p3 galaxies is an important mechanism for transporting
metals in the low density IGM (voids).
     
\def\capfigp{ (Left) Ratio of the volume- and mass-weighted mean
  metallicities, $\langle Z/Z_\odot \rangle_V/\langle Z/Z_\odot
  \rangle_M$, as a function of redshift for the 64L05noRAD (solid),
  64L05p2-2 (dotted), 64L05p3b (dashed), and 64L05p2noLW+1
  (long-dashed) runs. This ratio is proportional to the volume filling
  factor of the metal enriched gas. The lines, from the bottom to the
  top, have an increasingly bursting SF mode. The SF bursts are
  stronger when the parameter combination
  $\epsilon_*\epsilon_{UV}$\fesc is large. (Right) Mass-weighted mean
  metallicity, $\langle Z/Z_\odot \rangle_M$, as a function of
  redshift.}
\placefig{
\begin{figure*}[thp]
\epsscale{1.0}
\plottwo{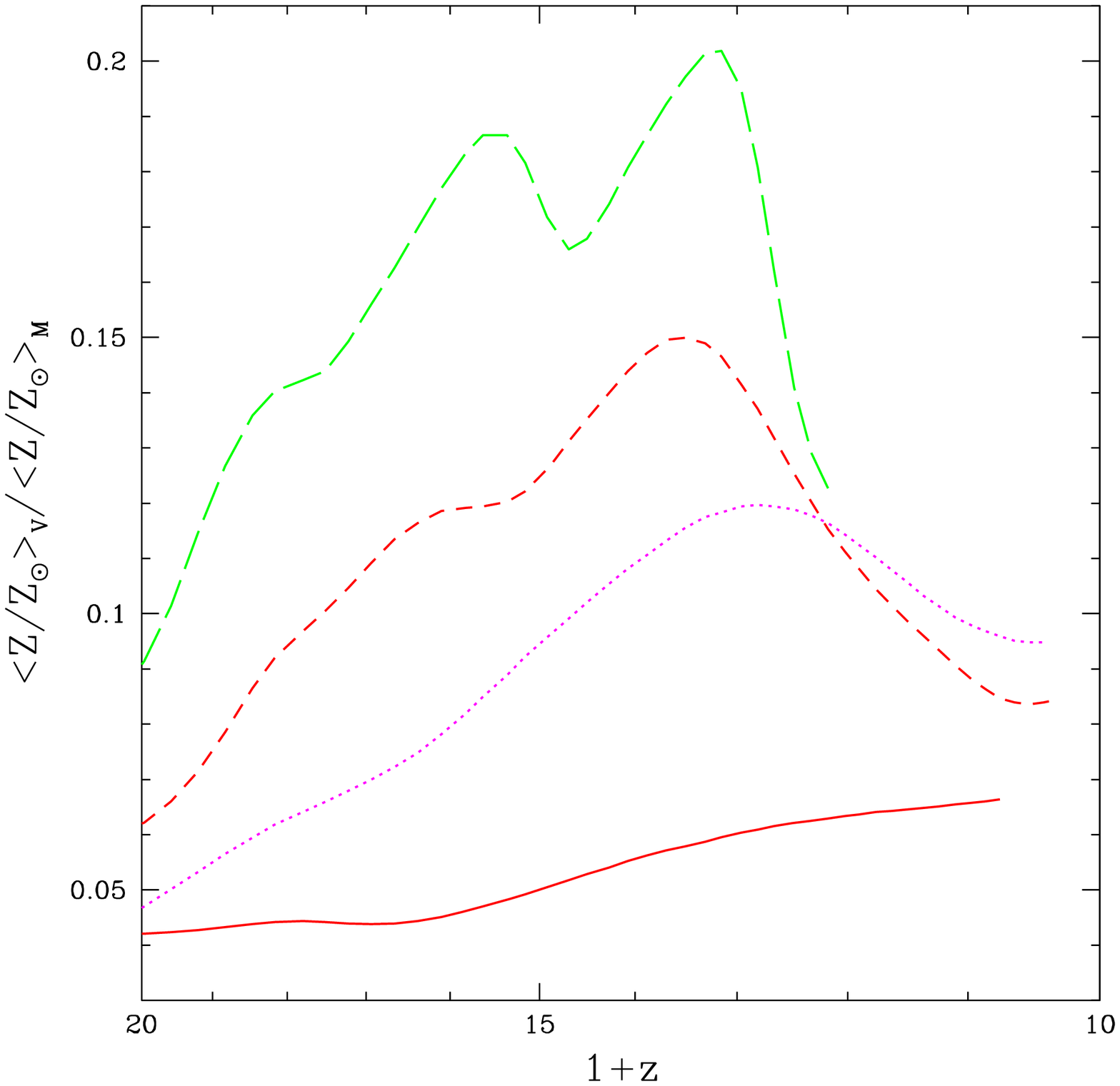}{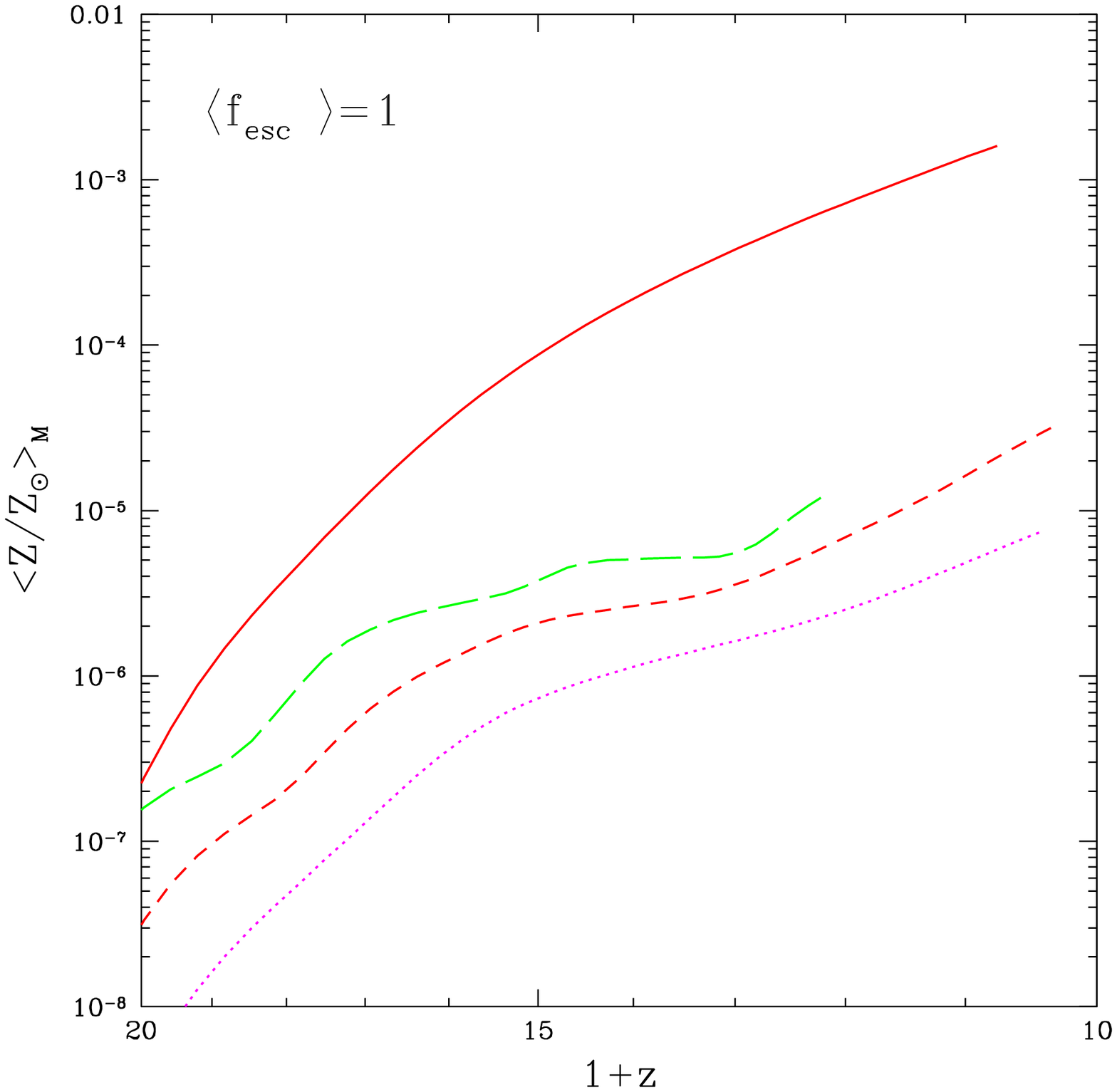}
\epsscale{1.0}
\caption{\label{fig:zt}\capfigp}
\end{figure*}
}

Figure~\ref{fig:zt}~(right) shows the mass-weighted metallicity
$\langle Z/Z_\odot \rangle_M$ for the four simulations in
Figure~\ref{fig:zt}~(left). It can be easily shown that $\langle
Z/Z_\odot \rangle_M \propto \epsilon_{UV} f_{star}$, where $f_{star}$
is the mass fraction in stars and $\epsilon_{UV}$ is the energy in
ionizing photons per rest mass energy of H atoms ($m_H c^2$)
transformed into stars (the number of ionizing photons emitted by a
population of stars is proportional to the number of heavy elements
released in the ISM). The parameter $\epsilon_{UV}$ depends on the IMF
and stellar metallicity. If the feedback of the EUV radiation
regulates the SFR, we have $f_{star} \propto (\epsilon_{UV} \langle
f_{esc}\rangle)^{-1}$. In this case, we expect $\langle Z/Z_\odot
\rangle_M \propto$\fesc\m, independent of the stellar metallicity and
IMF. We conclude that, while the SFR and $f_{stars}$ is inversely
proportional to the parameter combination $\epsilon_{UV}$\fesc, the
mass of metals produced by \p3 depends only on \fesc. In
Figure~\ref{fig:zt} we have used \fesc$=1$.

\subsection{Realistic Scenarios of Cosmic Evolution during the
  ``Dark Ages'' \label{ssec:realsim}}

In this section we show the evolution of the global properties of the
universe in our three most realistic simulations: the 128L1p2,
128L1p2-2 and 256L1p3 runs. The mass resolution ($M_{DM} =4.93\times 10^3$
\Ms for the $256^3$ and $3.94\times 10^4$ \Ms for the
$128^3$ runs) and box size, $L_{box}=1$ \Mpc, of these simulations is
sufficiently large to resolve the formation of the first \p3 and \pII
objects. In particular, the 128L1p2-2 and 256L1p3 simulations include
the effects of secondary electrons and the SED modification caused by
assuming realistic values of \fesc. We have \fesc$=0.1$,
$\epsilon_*=0.1$, and metal-free SED ($(\epsilon_{UV}/4\pi)=2.5 \times
10^{-5}$) for the 256L1p3 run. For the 128L1p2-2 run we assume
\fesc$=0.01$, $\epsilon_*=0.05$ and $Z=0.05 Z_\odot$ \popII SED
($(\epsilon_{UV}/4\pi)=1.1 \times 10^{-5}$). The 128L1p2 run has \fesc$=1$,
$\epsilon_*=0.2$, $(\epsilon_{UV}/4\pi)=1.6 \times 10^{-5}$ and \popII SED. We
believe that the $256^3$ run is very close to the limit of numerical
convergence.  Finally, the formal spatial resolution is $156$ \pc
comoving [$B_*=25$ is the parameter that regulates the maximum
deformation of the Lagrangian mesh: the spatial resolution is $\sim
L_{box}/(N_{box}B_*)$] in the 256L1p3 run, $488$ \pc comoving
($B_*=16$) in the 128L1p2-2 run, and $781$ \pc comoving ($B_*=10$) in
the 128L1p2 run. To give a better idea of the scales resolved by the
simulations, we remind the reader that the comoving core radius of a
just-virialized DM halo of mass $M_{DM}$ is $R_c \approx 300~{\rm
  pc}(M_{DM}/10^6 M_\odot)^{1/3}$ (we have assumed a halo
concentration parameter $c=R_{vir}/R_c=10$).  We refer the reader to
Paper~I for details on the physics included in the code and in our
convergence studies.

The thick solid lines in Figure~\ref{fig:sfr1} show the comoving SFR and
fraction of baryons in stars, $f_{star}$, as a function of redshift
for the 256L1p3 run (left panels), 128L1p2-2 run (center panels)
and 128L1p2 run (right panels).  As a comparison we plot the
same runs without radiative transfer (dashed lines) and the 64L1noC
run ($64^3$ cells and $L_{box}=1$ \Mpc), excluding both radiative
transfer and \H2 cooling (thin solid lines). Therefore, thin solid
lines show the contribution to the global SFR of \pII objects only,
and the dashed lines show the SFR of \p3 and \pII objects without any
feedback effect.  The main result shown in this figure is that \p3
objects are an extremely important (or dominant) fraction of the
galaxies at least until redshift $z \sim 10$.  Contrary to what is
widely believed, their formation is not severely suppressed by the
dissociating background. We showed in \S~\ref{ssec:LW} that the
dissociating background has little influence in determining the SF
history.  In \S~\ref{ssec:sim} we demonstrated that the mass
fraction of \p3 objects formed depends, instead, {\em only} on the
value of \fesc and the stellar IMF.

\def\capfigq{The thick solid lines show the SFR (top panels) and
  fraction of baryons in stars, $f_{star}$ (bottom panels), as a
  function of redshift for the 256L1p3 (left panels), 128L1p2-2
  (center panels) and 128L1p2 (right panels) simulations. The box size
  is 1 \Mpc comoving. The 256L1p3 run has a \pop3 SED, \fesc$=0.1$ and
  $\epsilon_*=0.1$; the 128L1p2-2 run has \popII SED, \fesc$=0.01$ and
  $\epsilon_*=0.05$, and the 128L1p2 run has \popII SED, \fesc$=1$ and
  $\epsilon_*=0.2$. The thin dashed lines show the same simulations,
  without radiative transfer, and the thin solid lines excluding both
  radiative transfer and \H2 cooling (\ie, the contribution of \pII
  objects).}
\placefig{
\begin{figure*}[thp]
\plotone{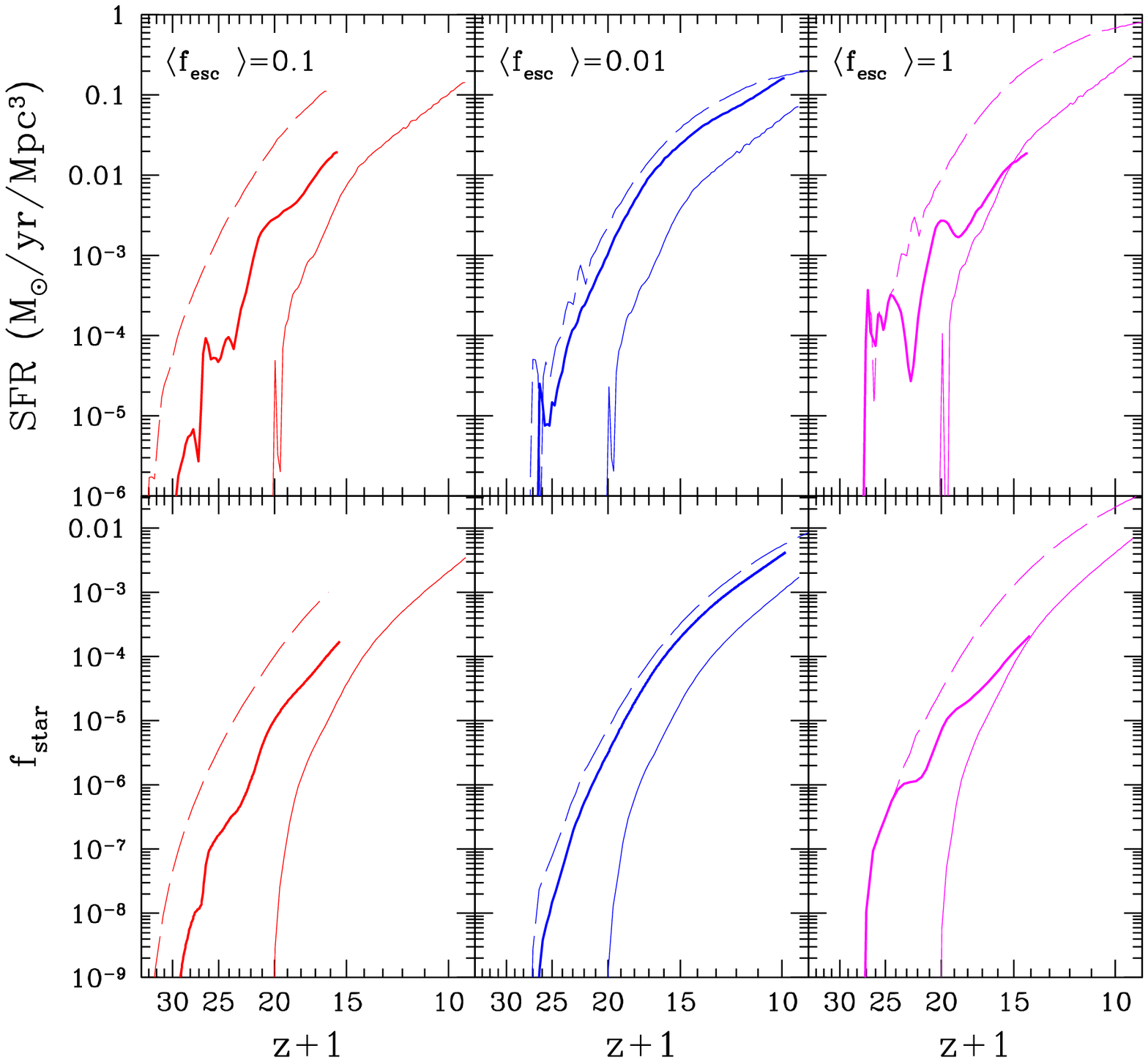}
\caption{\label{fig:sfr1}\capfigq}
\end{figure*}
}
 
The two panels on the left of Figure~\ref{fig:sfr1} show the SFR and
$f_{star}$ for the $256^3$ cells simulation which has \fesc$=0.1$; in
this simulation the first stars form at $z =30$.  At $z = 15$ the
fraction of stars in \p3 objects is $f_{star} \simeq 2 \times
10^{-4}$, about 5 times the mass of stars in \pII objects. The
radiative feedback has reduced $f_{star}$ by a factor of 10. Note that
$f_{star}$ for \pII objects (thin solid line) and for \p3 objects
without feedback (thin long-dashed line) scales with $\epsilon_*$.
The fraction of baryons transformed into stars of \p3 objects,
including radiative feedback effects, does not depend on $\epsilon_*$
as long as it is smaller than the the value without radiative
transfer. Since, for \pII objects, $f_{star} \propto \epsilon_*$, the
relative importance of \p3 to \pII objects is inversely proportional
to $\epsilon_*$ (which here is 0.1).  For instance, if we had chosen
$\epsilon_* < 0.1$ for this simulation, $f_{star}$ of \p3 objects
would have been as large as 50 times $f_{star}$ of \pII objects at $z
=15$, and the radiative feedback would have suppressed the SF by only
a small factor.

We show an example of such a case in the two central panels of
Figure~\ref{fig:sfr1}. In this $128^3$ simulation we use \fesc$=0.01$
and $\epsilon_*=0.05$. Here, the first stars form at $z =26$ because
of the lower mass resolution.  Radiative feedback suppresses the SFR
only by a factor of 2. At $z=15$, $f_{star} \simeq 3 \times 10^{-4}$,
and the contribution of \p3 objects to star formation is about 10
times the contribution of \pII objects. At $z=10$, \p3 objects still
dominate $f_{star}$ by a factor of 3.  Note that in this simulation
the dissociating background intensity is very high and, as shown by
the short-dashed lines in Figure~\ref{fig:xm1}, the H$^-$ and \H2
abundances in the IGM are extremely small ($\sim 10^{-12}$ relative to
H).

The two panels on the right show the SFR and $f_{star}$ for the
$128^3$ simulation, which has \fesc$=1$ and $\epsilon_* =0.2$. At $z =
15$, the fraction of stars in \p3 objects is $f_{star} \simeq 2 \times
10^{-5}$, about 1/5 the mass of stars in \pII objects. The radiative
feedback has reduced $f_{star}$ by a factor of 100.

In Figure~\ref{fig:xm1} we show the mass- and volume-weighted mean
molecular abundances $\langle x_{H_2} \rangle$, $\langle x_{H_2^+}
\rangle$ and $\langle x_{H^-} \rangle$ as a function of redshift for
the three simulations with radiative transfer shown in
Figure~\ref{fig:sfr1}.  The volume-weighted abundances are about two
orders of magnitude smaller than the mass-weighted abundance. This
shows that \H2 is much more abundant in dense regions. The
dissociating background destroys the \H2 in the low-density IGM in the
redshift range $20 < z <25$, depending on the choice of the free
parameters in the simulation. In the dense regions, the
production/destruction of \H2 sets its abundance to a quasi-constant
value that depends on the SED of the sources. At redshift $z \sim 15$,
the abundance of H$^-$, the main catalyst for \H2 production, reaches
its maximum value and starts to decrease.  Consequently, the \H2
abundance also decreases after $z \sim 15$.

\def\capfigr{ Mass- and volume-weighted molecular abundances $\langle
  x_{H_2} \rangle$, $\langle x_{H_2^+} \rangle$ and $\langle x_{H^-}
  \rangle$ for the three runs from Figure~\ref{fig:sfr1}. The solid
  lines show the 256L1p3 run (\fesc$=0.1$ and $\epsilon_* =0.1$), the
  short-dashed lines show the 128L1p2-2 run (\fesc$=0.01$ and
  $\epsilon_* =0.05$), and the long-dashed lines show the 128L1p2 run
  (\fesc$=1$ and $\epsilon_* =0.2$). The drops in \H2 abundance, while
  H$_2^+$ and H$^-$ are still high, are due to the rising flux of the
  dissociating radiation.}
\placefig{
\begin{figure*}[thp]
\plotone{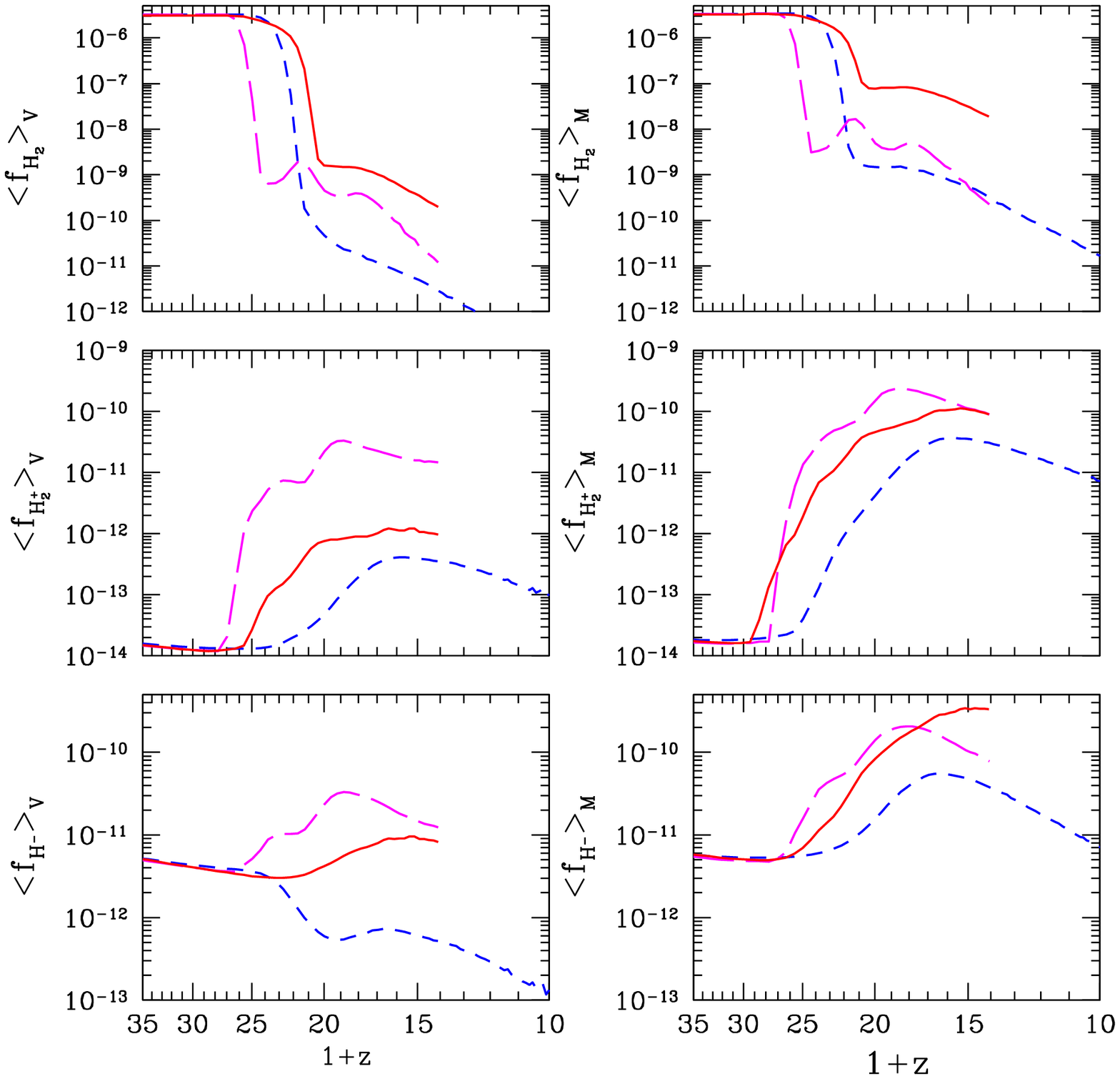}
\caption{\label{fig:xm1}\capfigr}
\end{figure*}
}

Figure~\ref{fig:xh1} is analogous to Figure~\ref{fig:xm1}, but here we
show the mass- and volume-weighted ionized H and He mean abundances
$\langle x_{\HII} \rangle$, $\langle x_{\GII} \rangle$, $\langle
x_{\GIII} \rangle$ (top), and the metallicity $\langle Z/Z_\odot
\rangle$ and temperature $\langle T \rangle$ (bottom). If reionization
takes place at redshift $z \sim 6$, as recent high-redshift quasar
observations seem to suggest \citep*{Becker:01, Djorgovski:01}, the
run shown by the long-dashed line reionizes the universe too early,
while the run shown by the short-dashed line reionizes the universe
too late. This observation does not constrain the value of \fesc or
$\epsilon_*$ for \p3 objects, since we expect that \fesc, at least,
will vary as a function of redshift. In this paper, we assume for
simplicity that \fesc, $\epsilon_*$, $\epsilon_{UV}$, and $g_\nu$ are
constants.

\def\capfigs{(Top) Mass- and volume-weighted ionized H and He
  abundances: $\langle x_{\HII} \rangle$, $\langle x_{\GII} \rangle$,
  $\langle x_{\GIII} \rangle$. (Bottom) Mass- and volume-weighted
  metallicity $\langle Z/Z_\odot \rangle$ and temperature $\langle T
  \rangle$. The solid lines show the 256L1p3 run (\fesc$=0.1$ and
  $\epsilon_* =0.1$), the short-dashed lines the 128L1p2-2 run
  (\fesc$=0.01$ and $\epsilon_* =0.05$), and the long-dashed lines show
  the 128L1p2 run (\fesc$=1$ and $\epsilon_* =0.2$).}
\placefig{
\begin{figure*}[thp]
\epsscale{0.6}
\plotone{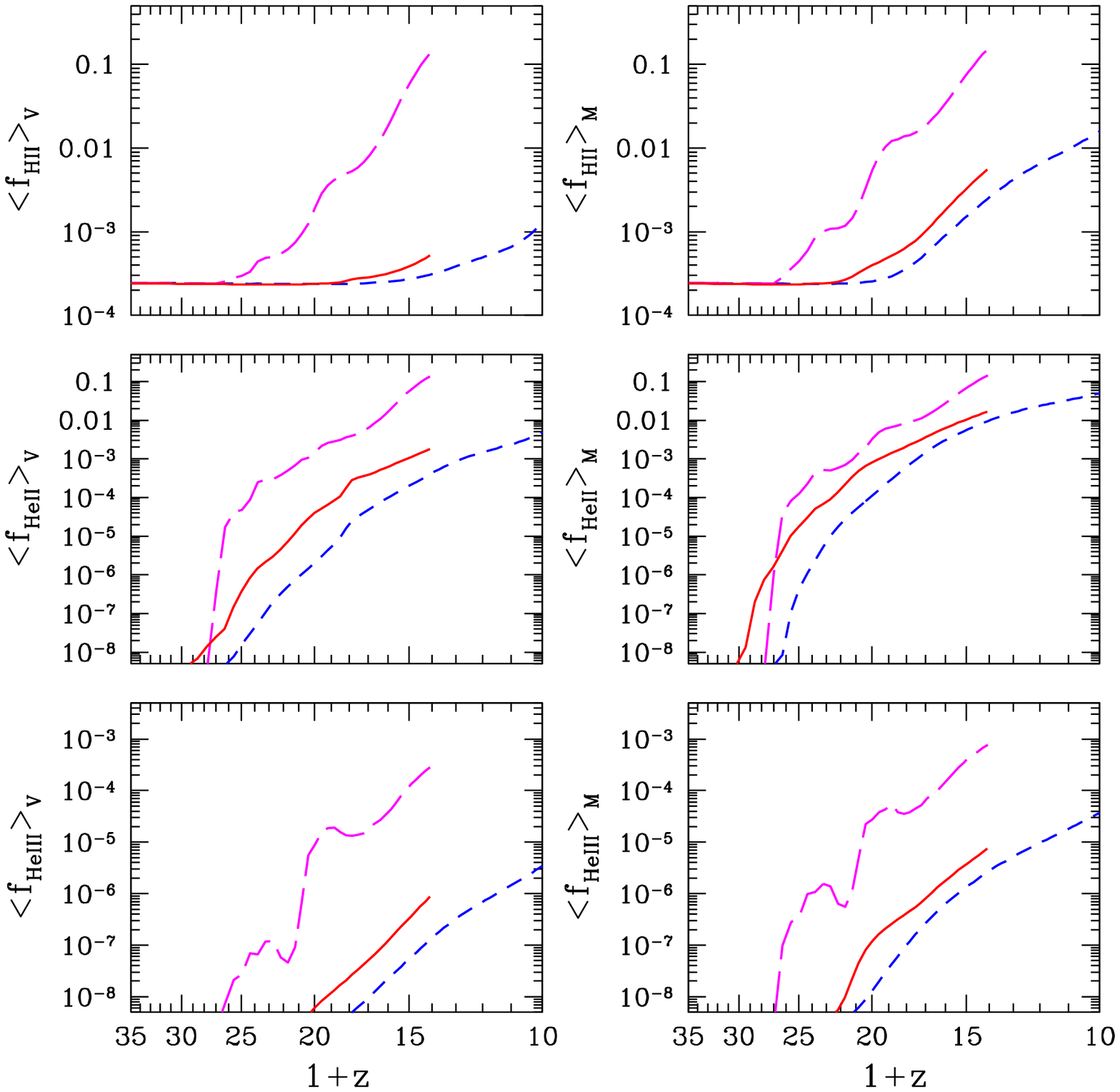}
\plotone{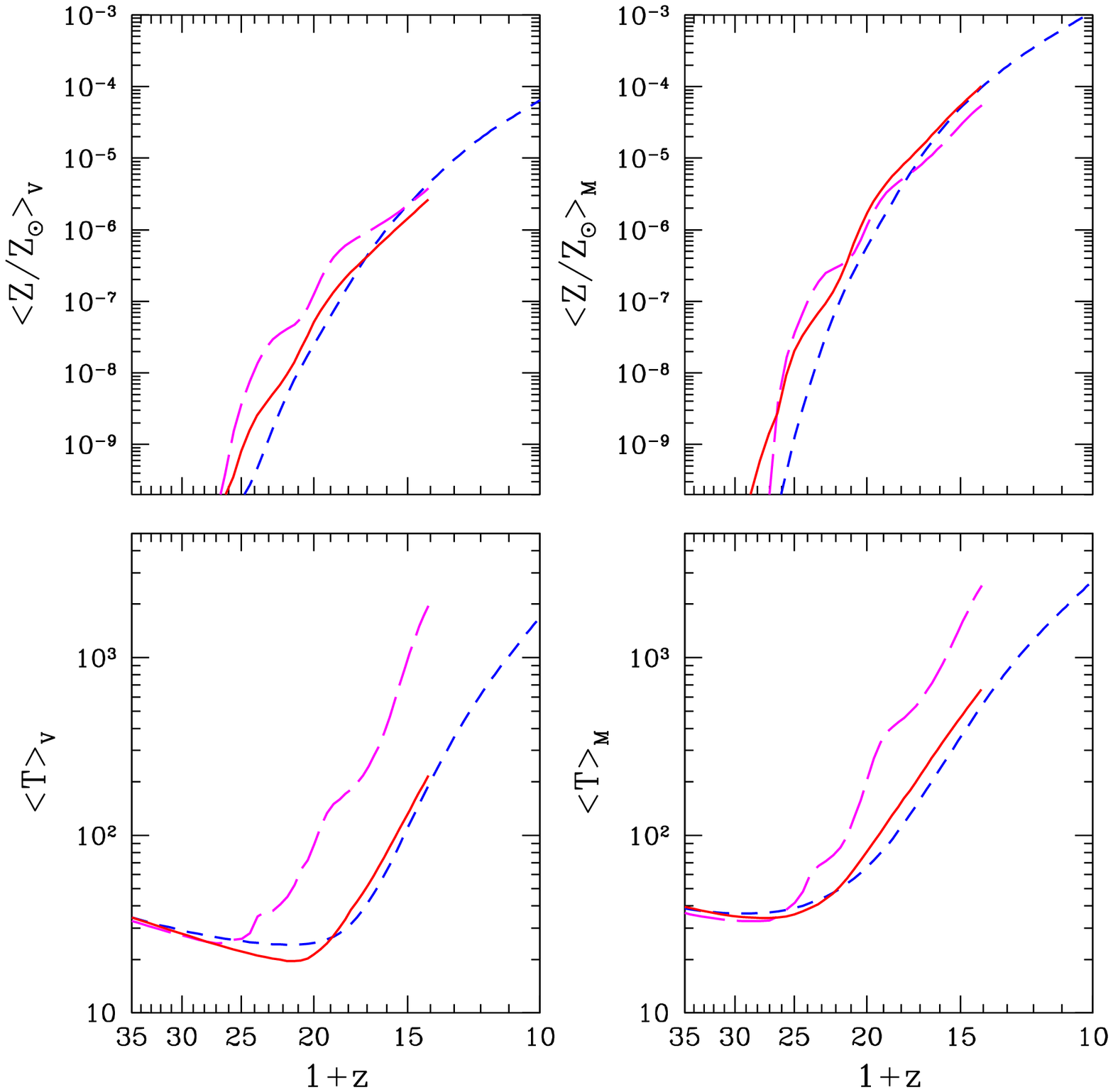}
\epsscale{1.0}
\caption{\label{fig:xh1}\capfigs}
\end{figure*}
}

The mass-weighted metallicity, $\langle Z/Z_\odot \rangle_M$, is
proportional to the total mass of metals produced by the stars, and
therefore is proportional to the SFR. Since we have shown that \p3
objects dominate the SFR if \fesc$<1$, at least before redshift $z
\sim 10$, their metal production dominates by the same amount.
Moreover, we have shown in \S~\ref{ssec:met} that, if SF is
bursting, photo-evaporation of small galaxies is the main process that
pollutes the low-density IGM (voids) with metals. Finally, SNe
explosions, not yet included in the simulations, can transport metals
into the voids efficiently, since \p3 objects are numerous and the
voids are small at high redshift.

In Figure~\ref{fig:j1}~(left) we show the evolution of the \HI
ionizing background $J_{\HI}$ (thick lines) and the \GII ionizing
background $J_{\GII}$ (thin lines) for the same three simulations
shown in the previous figure. In Figure~\ref{fig:j1}~(right) we show
the comoving mean free path of the \HI ionizing photons
$\lambda_{\HI}$ (thick lines) and of the \GII ionizing photons
$\lambda_{\GII}$ (thin lines).  Reionization, defined as the overlap
of \HII regions, occurs when $\lambda_{\HI} \sim D_s/2 \sim 0.1$ \Mpc,
where $D_s$ is the mean comoving distance between the ionizing sources
\citep{Gnedin:00}.

\def\capfigt{(Left) Evolution of the \HI ionizing background $J_{\HI}$
  (thick lines) and the \GII ionizing background $J_{\GII}$ (thin
  lines). (Right) Comoving mean free path of the \HI ionizing photons
  $\lambda_{\HI}$ (thick lines) and of the \GII ionizing photons
  $\lambda_{\GII}$ (thin lines). The solid lines show the 256L1p3 run
  (\fesc$=0.1$ and $\epsilon_* =0.1$), the short-dashed lines the
  128L1p2-2 run (\fesc$=0.01$ and $\epsilon_* =0.05$), and the
  long-dashed lines show the 128L1p2 run (\fesc$=1$ and $\epsilon_*
  =0.2$).}
\placefig{
\begin{figure*}[thp]
\plottwo{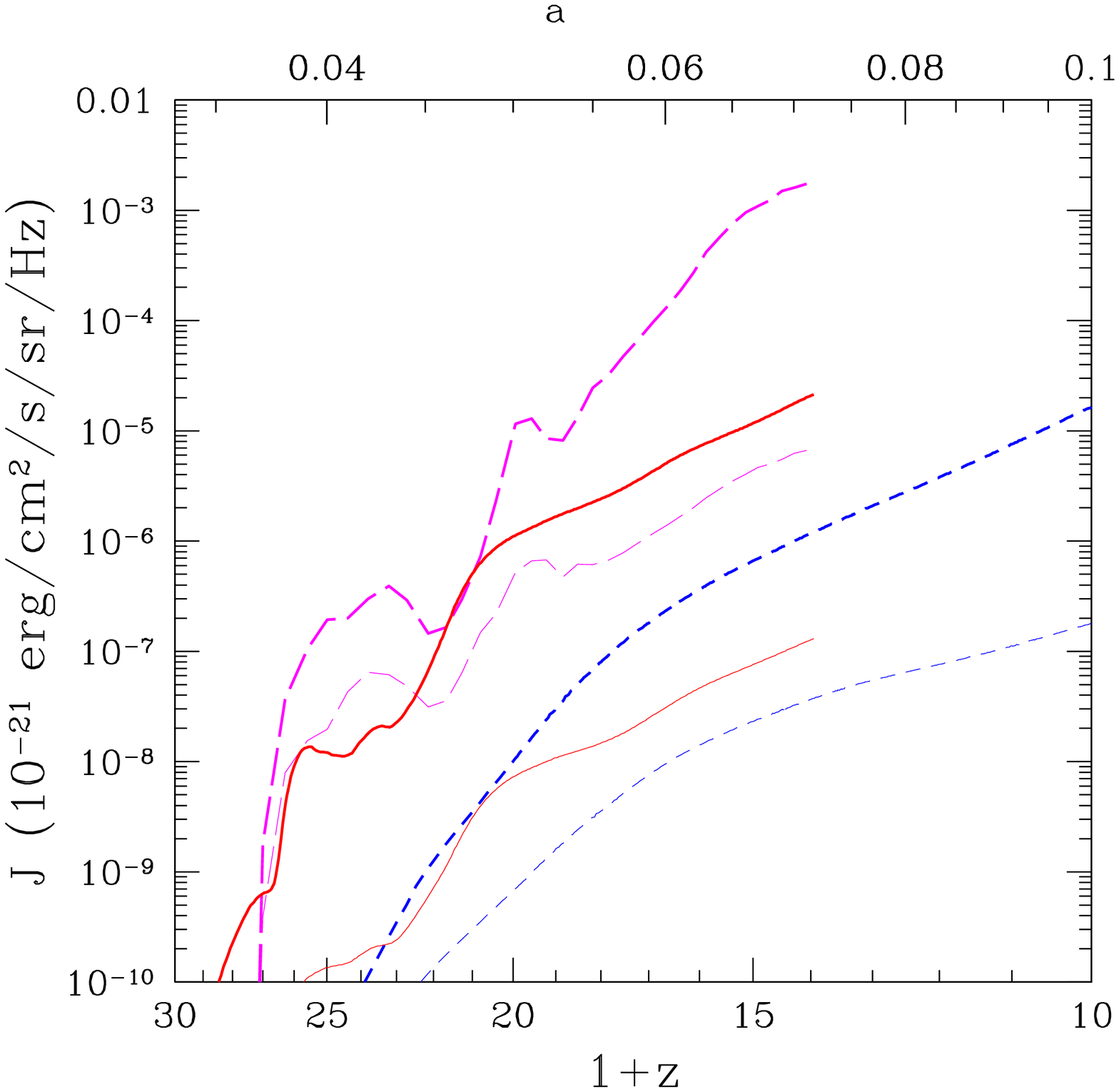}{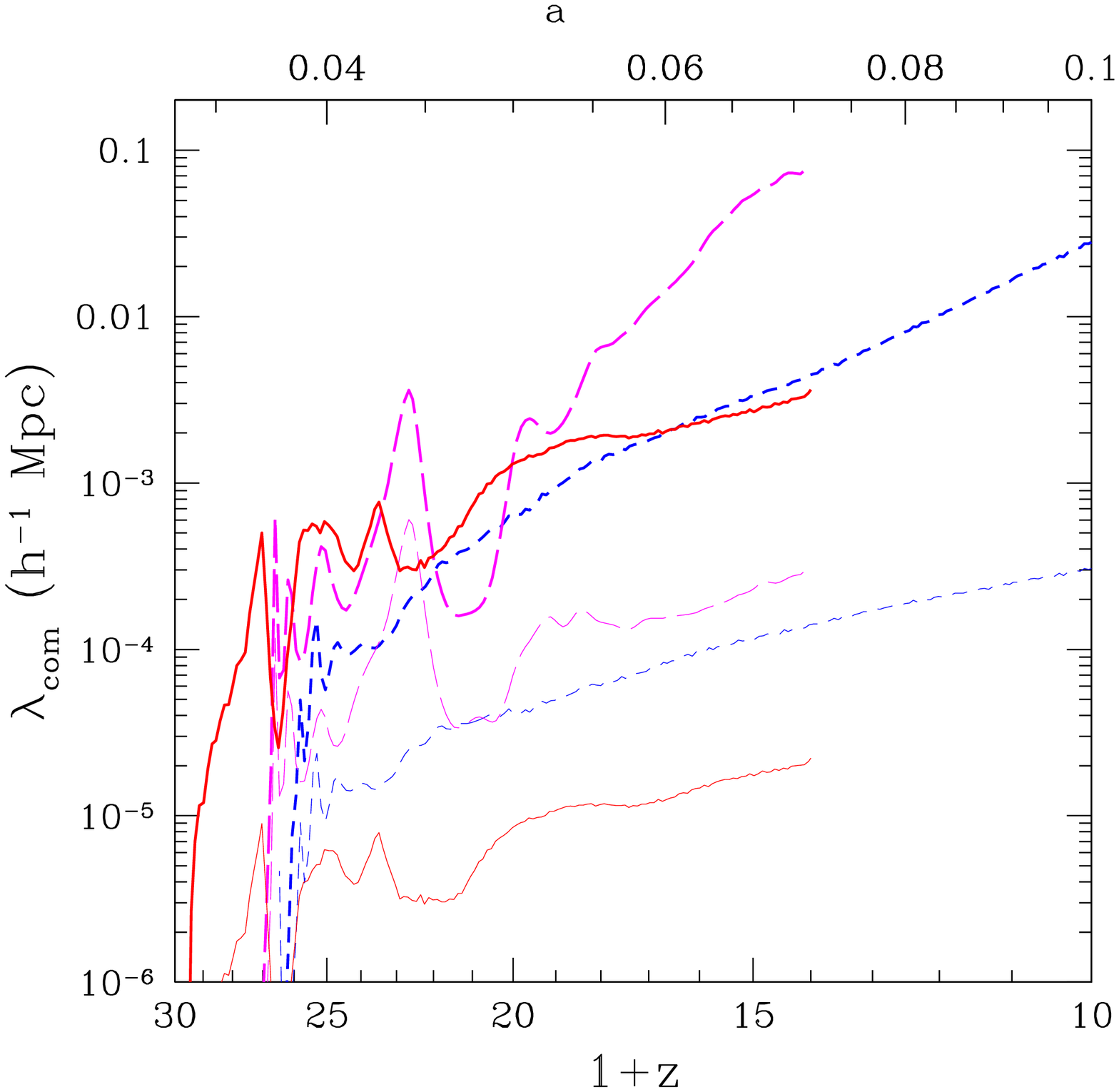}
\caption{\label{fig:j1}\capfigt}
\end{figure*}
}

\section{Discussion and Summary \label{sec:sum}}

In this paper we studied the the formation and evolution of the first
galaxies in a $\Lambda CDM$ cosmology. Our results are based on 3D
cosmological simulations that include, for the first time, a
self-consistent treatment of radiative transfer.  The simulations
include continuum radiative transfer using the ``Optically Thin
Variable Eddington Tensor'' (OTVET) approximation and line-radiative
transfer in the \H2 Lyman-Werner bands of the background radiation.
Chemical and thermal processes are treated in detail, particularly the
ones relevant for H$_2$ formation and destruction. The details about
the numerical methods and the physics included in the simulations are
treated in a companion paper (Paper~I). In Paper~I we have also
performed careful convergence analysis. It appears that we are very
close to the convergence limit when we use a $256^3$ box. In smaller
simulations the SF is underestimated.

The main result is that positive feedback processes dominate negative
feedback of the dissociating background, and therefore star formation
in \p3 objects is not suppressed. The main parameter that determine
the importance of \p3 objects is \fesc. If \fesc$\sim 0.01$, \p3
objects dominate the galaxy mass function at least until redshift $z
\sim 9$, and feedback produces a bursting SF in these objects. Because
the ionization fronts are confined to the dense filaments,
reionization of the IGM cannot be produced by \p3 objects. However, we
find that \p3 objects are important in enriching the low-density IGM
with heavy elements.

The main processes that are not included in our simulations are \H2
self-shielding and mechanical and thermal energy injection caused by
SN explosions. \H2 self-shielding, depending on the choice of the free
parameters, could be relevant and should reduce the effects of the
dissociating radiation. Since we have found that the
dissociating radiation has a negligible role in regulating SF, we
expect that the results of our simulations should not be modified by
\H2 self-shielding. The effects of dissociating radiation become
important when \fesc$\simlt 10^{-3}$. \H2 self-shielding effects
probably reduce the value of \fesc for which dissociating radiation
becomes important to a smaller value.  Although we have not included
them in our simulations, SN explosions can be important; their
feedback could be responsible for a self-regulating global SF and
contribute to spread metals into the low-density IGM.  Unfortunately,
the dynamical and thermal effects of SN explosions on the ISM of
galaxies and IGM are not fully understood.  The investigation of the
effects of mechanical feedback from SNe explosions will be the subject
of our future work.

We cannot rule out alternative cosmologies, in which \p3 objects do
not form.  For example, in warm dark matter (WDM) cosmologies, the
free-streaming of the DM particles is large enough to suppress the
formation of dwarf galaxies at high redshift. It is also possible that
the initial power spectrum does not have enough power on small scales
to form \p3 objects. Theoretical arguments based on observations of
the Ly$\alpha$ forest, quasars at high redshift, and stellar
populations in dwarf galaxies pose some constraints on the mass of WDM
particle candidates ($m_{wdm} \simgt 1$ keV).  Current observations do
not rule out these alternative cosmological models; but they may not
solve the problems faced by CDM.

In the following list we summarize the results of this work in
more detail:

\noindent
(1) {\bf SFR in \p3 objects is self regulated and depends only on
  \fesc,} for fixed $\epsilon_{UV}$ (known IMF). The SFR is almost
independent of the star formation efficiency, $\epsilon_*$, the
dissociating background, and the SED (metal-free or metal-rich
objects) of the sources. It depends only on \fesc and the IMF (number
of ionizing photons per baryon converted into stars). If \fesc is
small, the SFR is high, while if \fesc$\sim 0.01$ the SF is only
slightly suppressed by radiative feedback. In this case, the maximum
SFR is proportional to $\epsilon_*$, and \p3 objects dominate the
galaxy mass function at least until redshift $z \sim 10$.

\noindent
(2) {\bf \pp3 star formation is intrinsically ``bursting''.} Star
formation is regulated by competing negative and positive feedback
from EUV (ionizing) radiation.  The \HII regions produced by \p3
objects are confined to the dense IGM filaments. Reionization of the
IGM cannot happen until massive galaxies are formed.  In contrast to
massive objects, which reionize voids first, \p3 objects partially
ionize only the dense filaments while leaving the voids neutral.

\noindent
(3) {\bf Galaxy formation is triggered by the presence of neighboring
  galaxies.}  The SFR does not depend strongly
on the dissociating background intensity or on the H$_2$ abundance in the
IGM. The self-regulation of star formation relies on H$_2$
being continuously reformed in positive feedback regions. The H$_2$
formation happens only in the filaments, both inside relic \HII
regions that never expand into the low density IGM and in the H$_2$
shells just in front of the \HII regions \citep{RicottiGS:01}.

\noindent
(4) {\bf The dissociating (FUV) radiation reduces the SFR only
  if \fesc is very small.} If \fesc$\simlt
10^{-3}$ for a \popII SED or \fesc$\simlt 10^{-4}$ for a \pop3 SED, the
negative feedback of the dissociating background suppresses \p3
object formation, more efficiently at high-redshift.

\noindent
(5) {\bf \pp3 objects dominate the metal pollution of the low-density
  IGM.} The transport of metals from the galaxies to the low-density
IGM happens because of the continuous formation and photo-evaporation
of small mass (\p3) halos. The metal production, if the SF is
self-regulated by EUV radiation, is independent of the SFE, SED, and
IMF, but depends only on \fesc.
    
In conclusion, we have shown that, if SN explosions do not suppress
the formation of \p3 objects and CDM cosmogonies prove to be correct,
\p3 objects should have profound effects on cosmic evolution.
Observations of dwarf spheroidal galaxies, metallicity of the
Ly$\alpha$ forest, and stellar populations in the halo of the Milky
Way could verify this model. Computational limitations prevent us from
evolving a representative sample of the universe to redshifts $z < 9$.
At lower redshifts, the bulk of \p3 objects merge, forming larger mass
galaxies, but some of them might survive almost unaffected by the
environment.  Reionization is probably affecting the ISM of these
small galaxies quite substantially, photoevaporating the remaining
unshielded gas. These ``fossil'' \p3 objects could be identified with
at least some dwarf spheroidal galaxies in the Local Group.  In a
paper currently in preparation, we study the properties of the
simulated \p3 objects in order to understand whether this link is
real.  The prediction of a large population of \p3 objects offers a
challenging test to verify CDM cosmogonies.  We speculate that, in the
near future, given the rapid progress of computational power and the
rapid growth of observational data of cosmological interest, detailed
cosmological simulations will allow us to constrain the evolution of
our free-parameters: \fesc, $\epsilon_{UV}$, and $\epsilon_*$, or
possibly the nature of the dark matter.

\acknowledgements 
  
This work was supported by the Theoretical Astrophysics program at the
University of Colorado (NASA grant NAG5-7262). The simulations
presented in this paper were performed using SGI/CRAY Origin 2000
array at the National Center for Supercomputing Applications (NCSA).
Massimo Ricotti is grateful to Erika Yoshino for helping to make the
first draft of the manuscript readable.

\clearpage
\bibliographystyle{/home/origins/ricotti/Latex/TeX/apj}
\bibliography{/home/origins/ricotti/Latex/TeX/archive}

\vskip 2truecm

\placefig{\end{document}}

\clearpage

\newcounter{figurecap}
\setcounter{figurecap}{0}

\begin{center}
\bf Figure Captions
\end{center}

\refstepcounter{figurecap}
Fig.\ \thefigurecap---\label{fig:r1}\capfiga

\refstepcounter{figurecap}
Fig.\ \thefigurecap---\label{fig:tile}\capfigb

\refstepcounter{figurecap}
Fig.\ \thefigurecap---\label{fig:tile1}\capfigc

\refstepcounter{figurecap}
Fig.\ \thefigurecap---\label{fig:r2}\capfigd

\refstepcounter{figurecap}
Fig.\ \thefigurecap---\label{fig:radLW}\capfige

\refstepcounter{figurecap}
Fig.\ \thefigurecap---\label{fig:slice1}\capfigf

\refstepcounter{figurecap}
Fig.\ \thefigurecap---\label{fig:slice2}\capfigg

\refstepcounter{figurecap}
Fig.\ \thefigurecap---\label{fig:r3}\capfigh

\refstepcounter{figurecap}
Fig.\ \thefigurecap---\label{fig:r4}\capfigi

\refstepcounter{figurecap}
Fig.\ \thefigurecap---\label{fig:fesc}\capfigl

\refstepcounter{figurecap}
Fig.\ \thefigurecap---\label{fig:fesc3}\capfigm

\refstepcounter{figurecap}
Fig.\ \thefigurecap---\label{fig:ske}\capfign

\refstepcounter{figurecap}
Fig.\ \thefigurecap---\label{fig:zrend}\capfigo

\refstepcounter{figurecap}
Fig.\ \thefigurecap---\label{fig:zt}\capfigp

\refstepcounter{figurecap}
Fig.\ \thefigurecap---\label{fig:sfr1}\capfigq

\refstepcounter{figurecap}
Fig.\ \thefigurecap---\label{fig:xm1}\capfigr

\refstepcounter{figurecap}
Fig.\ \thefigurecap---\label{fig:xh1}\capfigs

\refstepcounter{figurecap}
Fig.\ \thefigurecap---\label{fig:j1}\capfigt

\clearpage

\tabone

\end{document}